\documentclass[]{spie}  %>>> use for US letter paper
\def \x     {{\times}}
\newcommand{\norm}[1]{\left\|#1\right\|}
\usepackage{subfigure}
\newcommand{\mF}{{\mathcal{F}}}
\newcommand{\bw}{{\mathbf{w}}}
\newcommand{\bK}{{\mathbf{K}}}
\newcommand{\bS}{{\mathbf{S}}}
\newcommand{\bd}{{\mathbf{d}}}
\newcommand{\wh}[1]{\widehat{#1}}
\newcommand{\abs}[1]{\left|#1\right|}
\newcommand{\set}[1]{\big\{#1\big\}}
\newcommand{\ceil}[1]{\left\lceil #1 \right\rceil}
\newcommand{\be}{\begin{equation}}
\newcommand{\ee}{\end{equation}}
\newcommand{\bm}{{\mathbf{m}}}
\newcommand{\mS}{{{S}}}
\newcommand{\mW}{{{A}}}
\newcommand{\cD}{{{\mathcal D}}}
\newcommand{\cR}{{{\mathcal R}}}
\newcommand{\mT}{{\mathcal{T}}}
\newcommand{\bnabla}{\mbox{\boldmath$\nabla$}}
\usepackage[]{graphicx,bbold}
\usepackage[]{subeqn,url}
\title{Wavelets and wavelet-like transforms on the sphere\\
and their application to geophysical data inversion}
\author{
Frederik J.~Simons\supit{a}, 
Ignace Loris\supit{b},
Eugene Brevdo\supit{c},
Ingrid C. Daubechies\supit{d},
\skiplinehalf
\supit{a}
Department of Geosciences, Princeton University, Guyot Hall,
Princeton, NJ, USA\\
\supit{b}
Department of Mathematics, Universit\'e Libre de Bruxelles, Belgium\\
\supit{c}
Department of Electrical Engineering, Princeton University, Princeton, NJ, USA\\
\supit{d}
Department of Mathematics, Duke University, Durham, NC, USA}
\begin{document} 
\maketitle 

%%%%%%%%%%%%%%%%%%%%%%%%%%%%%%%%%%%%%%%%%%%%%%%%%%%%%%%%%%%%% 
\begin{abstract}
  Many flexible parameterizations exist to represent data on the
  sphere. In addition to the venerable spherical harmonics, we have
  the Slepian basis, harmonic splines, wavelets and wavelet-like
  Slepian frames. In this paper we focus on the latter two: spherical
  wavelets developed for geophysical applications on the cubed sphere,
  and the Slepian ``tree'', a new construction that combines a
  quadratic concentration measure with wavelet-like multiresolution.
  We discuss the basic features of these mathematical tools, and
  illustrate their applicability in parameterizing large-scale global
  geophysical (inverse) problems.
\end{abstract}


\section{Introduction}
\label{intro}

This paper is about parameterization, and its role in geophysical
inverse problems: the analysis and representation of volumetric
properties, with a particular emphasis on the three-dimensional ball
and its surface, the two-dimensional sphere. This is not the sole
purview of geophysics (e.g. geodesy, geodynamics, seismology): 
there is a large literature on the subject in virtually every area of
scientific inquiry (e.g. medical imaging, astronomy, cosmology,
computer graphics, image processing,~...). For this reason we limit
ourselves here to a few technical aspects, some of which have been
published before (refs~\citenum{Simons+2011b}, \citenum{Brevdo2011})
but are illustrated with different examples.  

We begin with a new class of spherical wavelet basis on the ball
designed for the analysis of geophysical models and for the
tomographic inversion of global seismic
data\cite{Loris+2007,Loris+2010,Simons+2011b}. Its multiresolution
character allows for modeling with an effective spatial resolution
that varies with position within the Earth. We introduce two types of
discrete wavelet transforms in the angular dimension of the ``cubed
sphere'', a well-known Cartesian-to-spherical
mapping\cite{Ronchi+96,Komatitsch+2002a}.  These are applied to
analyze the information of terrestrial topography data and in seismic
wavespeed models of the Earth's mantle across scale space. The
localization and sparsity properties of the wavelet bases
allow finding a sparse solution to inverse problems by iterative
minimization of a combination of the $\ell_2$~norm of the data
residuals and the $\ell_1$~norm of the model wavelet coefficients.
These have now been validated in realistic synthetic experiments to
likely yield important gains in the inversion of seismic data for
global seismic tomography, the procedure by which seismic waveforms
are being used to image the three-dimensional distribution of elastic
(compressional or \textit{P} and shear or \textit{S}) wave speeds
($V_P$ and $V_S$) inside the Earth.

A second class of transforms is inspired by the type known as Slepian
functions. These we understand to be families of orthogonal functions
that are all defined on a common, e.g. geographical, domain, where
they are either optimally quadratically concentrated or within which
they are exactly limited, and which at the same time are exactly
confined within a certain bandwidth, or maximally concentrated
therein. Originally developed for the study of time series by Slepian,
Landau and Pollak\cite{Slepian+61,Landau+61} in the 1960s, they have
now been fully extended to the multidimensional Cartesian
plane\cite{Slepian64,Simons+2011a} and the surface of the
sphere\cite{Albertella+99,Wieczorek+2005,Simons+2006a}, where most of
the geophysical applications lie. We have reported on some aspects of
spherical Slepian functions and their role in the analysis and
representation of geophysical data in previous papers in this
series\cite{Simons+2007,Simons+2009b}. In this contribution we
announce a hybrid construction that blends desirable aspects and
properties of Slepian functions with ideas from wavelet and frame
theory. We summarize the main ideas contained in Chapter~7,
\textit{Multiscale Dictionaries of Slepian Functions on the Sphere},
of the Ph.~D. thesis by Eugene Brevdo\cite{Brevdo2011}, deferring a
full publication to a later date.

\newpage

\section{Wavelets on the Cubed Sphere}
\label{firstcon}

The use of wavelets is still no matter of routine in global geophysics,
beyond applications in one and two Cartesian dimensions. This despite
there being a wealth of available constructions on the
sphere\cite{Schroder+95,Narcowich+96,Antoine+2002,Holschneider+2003,Freeden+2004a,Fernandez+2006,Hemmat+2005,Schmidt+2006,Starck+2006,
  McEwen+2007,Wiaux+2007,Lessig+2008,Bauer+2011}. However, the cited
studies are mostly concerned with surfaces, not volumes, and many of
them require custom design and special algorithms. For the
application to global seismic tomography that we envisage, we take
advantage of the ease and flexibility of existing Cartesian
constructions by implementing standard algorithms on a
spherical-to-Cartesian map proposed by Ronchi et al.~(1996), a grid that
has long since proven its utility in the geosciences and beyond. 

\begin{figure}\center
\rotatebox{0}{
\includegraphics[width=0.4\columnwidth]{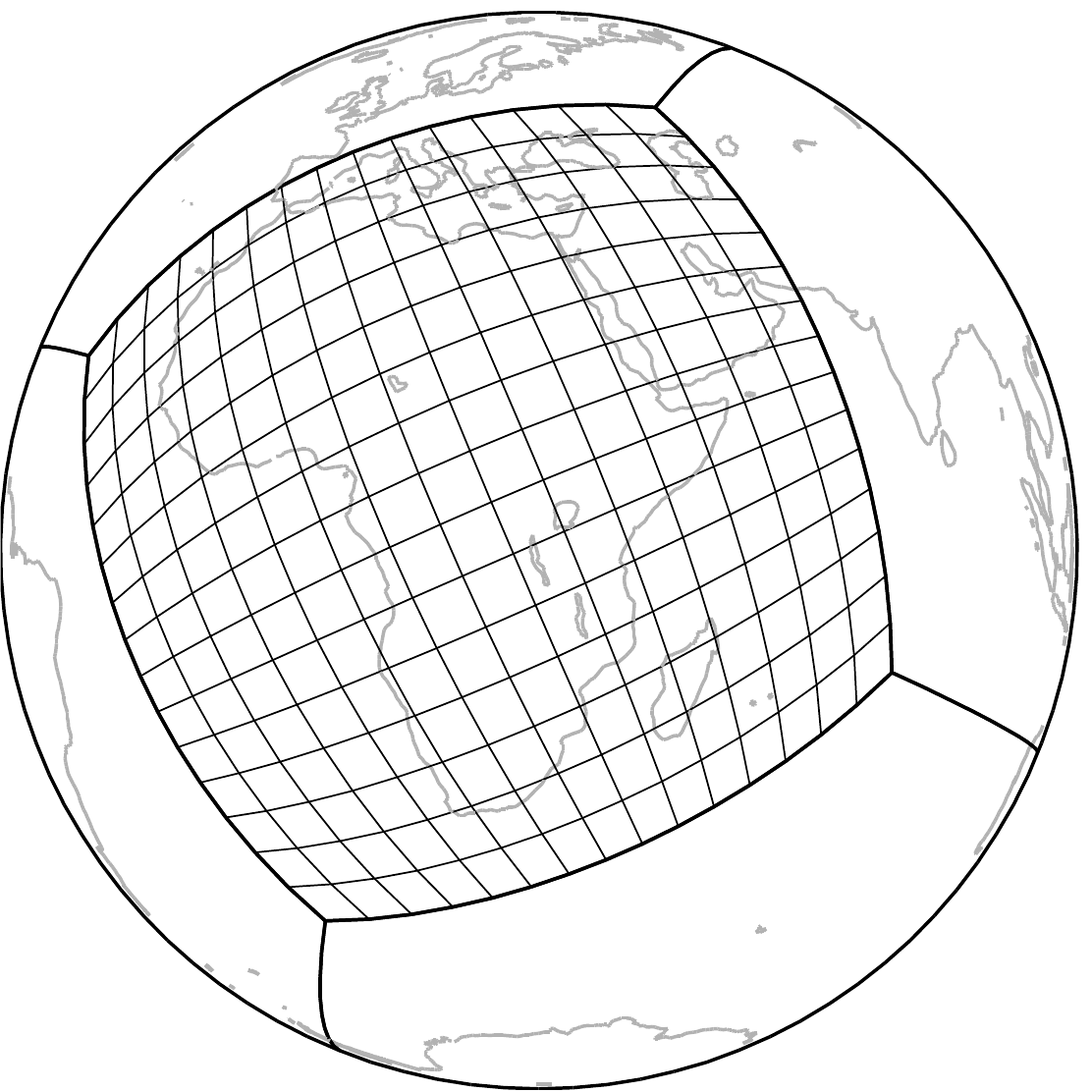}
\hspace{4em}
\includegraphics[width=0.4\columnwidth]{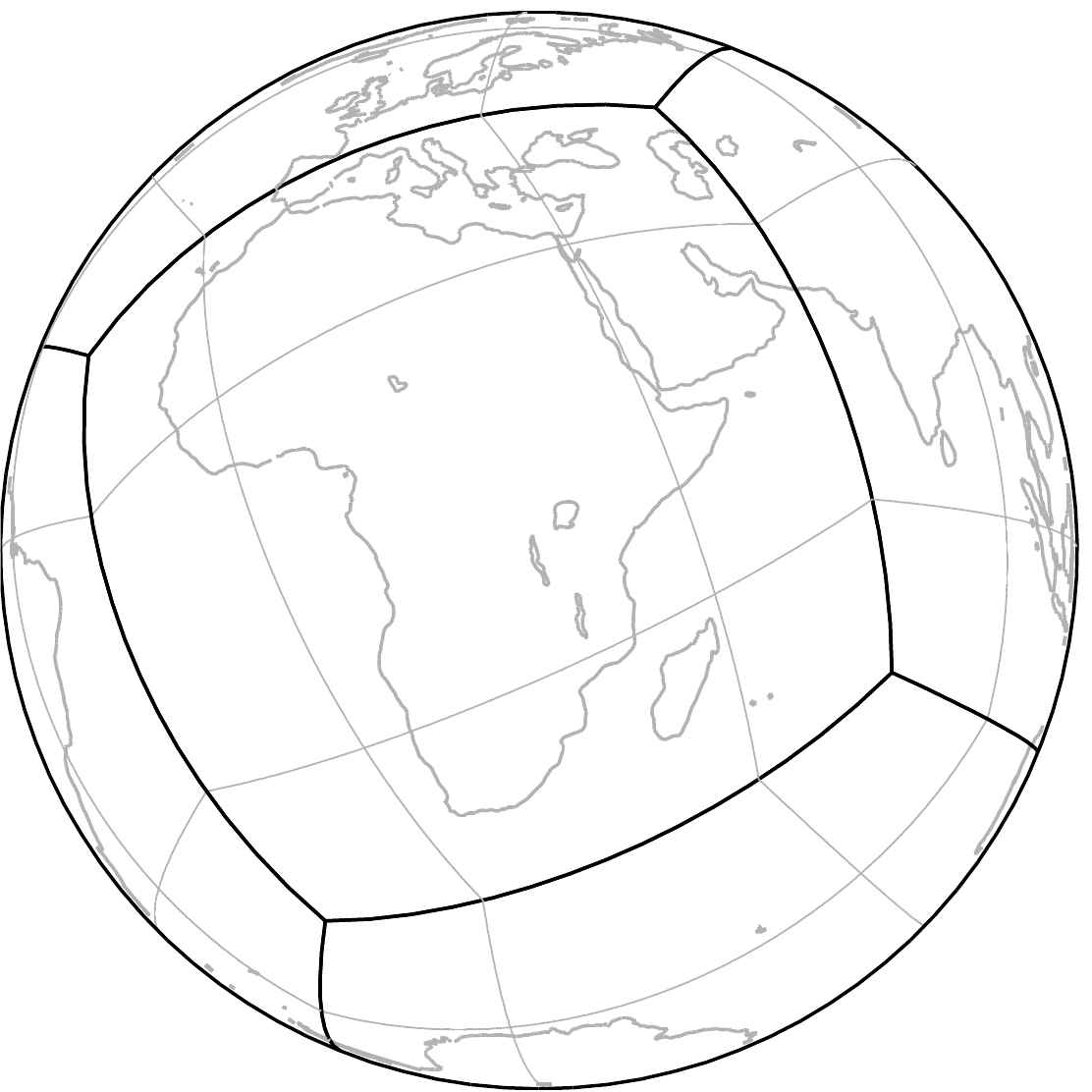}
}
\caption{Aerial view showing our adaptations of
the cubed sphere of ref.~\citenum{Ronchi+96}. Of the front-facing four of
the in total six ``chunks'', one is gridded to reveal its
$2^{2N}$ surface elements ($N=4$). The thick black lines identify
the boundaries of the six ``chunks''. The thin gray lines correspond to the
boundaries of the overlapping ``superchunks'' as discussed in the text.}
\label{thesuperchunk}\label{thechunk}
\end{figure}

As in ref.~\citenum{Ronchi+96}, we define a coordinate
4-tuple $(\xi,\eta,r,\kappa)$ for each of the $\kappa=1\rightarrow
6$ ``chunks'' comprising this ``cubed sphere'' at radius~$r$. In
ref.~\citenum{Simons+2011b} we list explicit formulas for the forward
and inverse mapping of cubed-sphere to Cartesian coordinates. Here we
suffice to say that a ``master'' surface chunk defined by the $2\times
2^N$ linearly spaced coordinate pairs
$-\pi/4\le(\xi,\eta)\le\pi/4$ % and the Jacobian
% \be
% J(\xi,\eta)=(1+\tan^2\!\xi)(1+\tan^2\!\eta)(1+\tan^2\xi+\tan^2\eta)^{-3/2}
% \ee
maps, in a general sense, to the Cartesian coordinate vector
\be
[\,x,y,z\,]=[1 \quad \tan\eta \quad -\tan\xi\,]\,(1+\tan^2\xi+\tan^2\eta)^{-1/2}
,
\ee
which is rotated to occupy a total of six non-overlapping patches tiling the
sphere, after which the entire construction is rotated over a standard
set of Euler angles $\alpha=0.0339$, $\beta=1.1705$, and
$\gamma=1.1909$. This results in the configuration shown in
Figure~\ref{thechunk} (\textit{left}). 

\begin{figure*}
\includegraphics[width=0.9\textwidth]{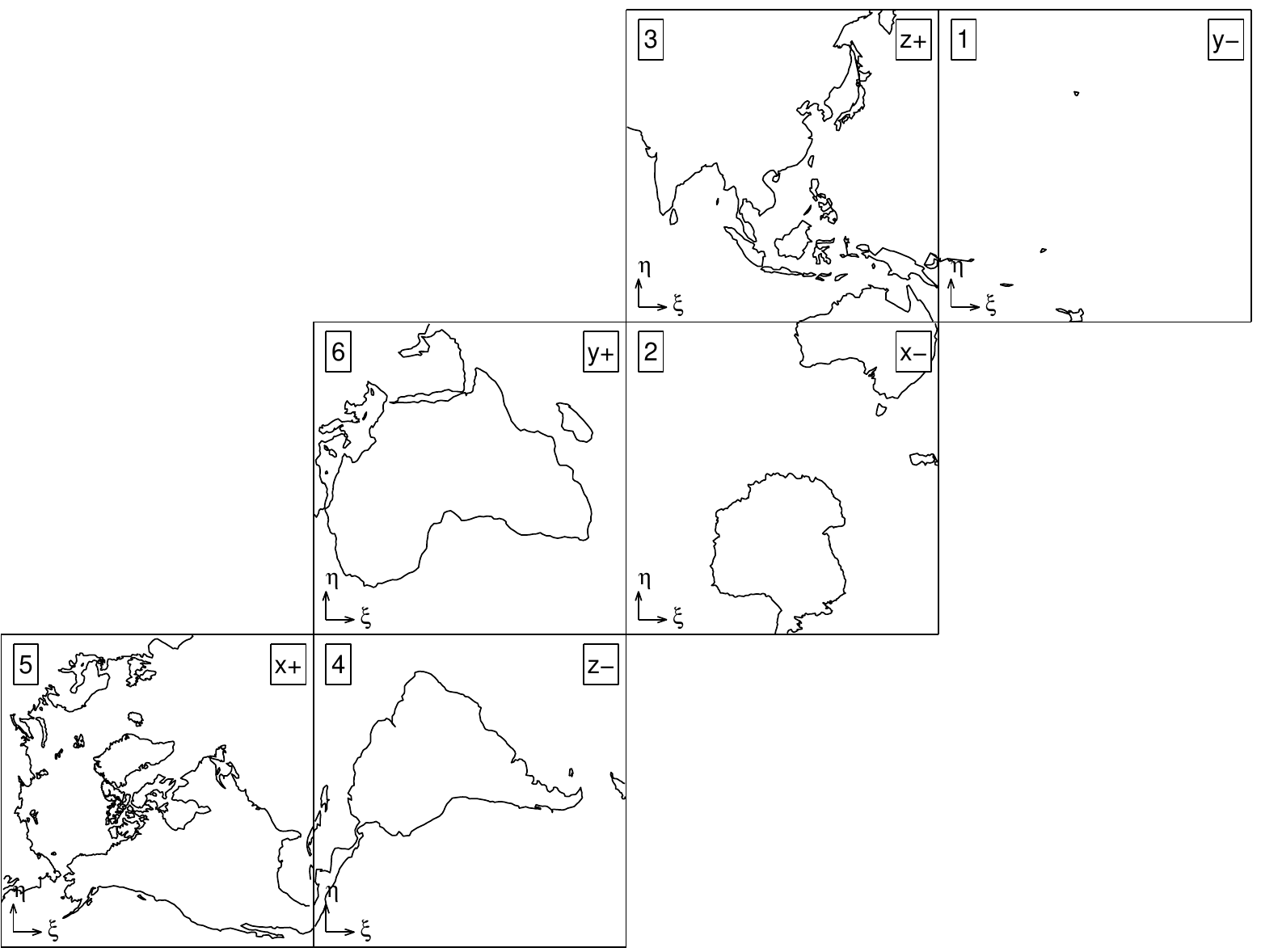}
\caption{\label{twodtopo}Geometry, nomenclature, and numbering of the
six faces of our cubed sphere in a two-dimensional view.}\vspace{-1em} 
\end{figure*}

Alternatively, for reasons that will be made clear, we
start from the coordinates $-3\pi/8\le(\xi,\eta)\le3\pi/8$ and
apply the same rotations, thereby producing six
overlapping ``superchunks'', as in Figure~\ref{thechunk}
(\textit{right}). Throughout this paper we will quote $N$ as the
angular resolution level of our cubed sphere, which implies that it
has~$6\times 2^{2N}$ such elements, with typical seismic tomography
grids having $N=7$. The Euler angles used in our construction were
chosen for geographical convenience, as can be seen from the
continental outlines in Figure~\ref{twodtopo}.

On these two grids we apply standard Cartesian transforms,
e.g. orthogonal\cite{Daubechies88b} and bi-orthogonal\cite{Cohen+92} ones,
whereby for the case of the non-overlapping ``chunks'' construction we
accommodate the presence of the ``seams' by switching to special
boundary filters at each of the edges, and applying preconditioners to
the data prior to  transformation in order to guarantee the usual
polynomial cancellation throughout the closed rectangular interval, as
is well established\cite{Cohen+93}. With this choice of bases,
sparsity in the representation of many data types is to be generally
expected\cite{Mallat2008}. 

\begin{figure*}\centering
\includegraphics[width=0.85\textwidth]{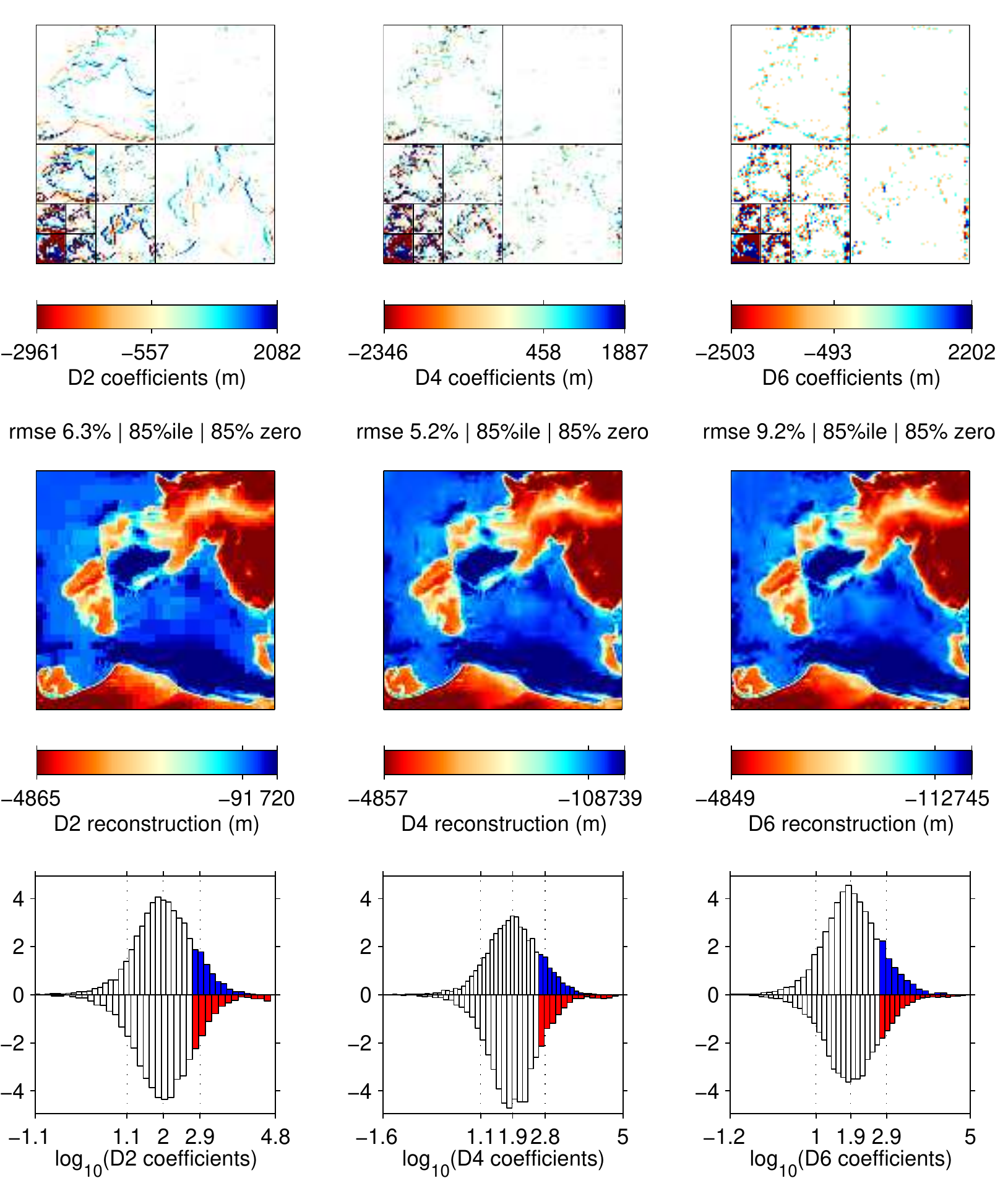} 
\caption{\label{wavelettopo} Wavelet and scaling coefficients
  (\textit{top}), space-domain reconstructions after thresholding
  (\textit{middle}), and ``signed'' histograms (\textit{bottom}) of
  the wavelet and scaling coefficients of the ``North American'' face
  of the cubed-sphere version of the Earth's topography, to a $2^J$
  dyadic subdivision with $J=3$. We used preconditioned interval
  wavelet transforms. All coefficients were hard-thresholded,
  retaining only the 15\% largest coefficients by absolute value. In
  the top row, the locations of zeroed coefficients are rendered
  white; those are also captured by the white bars in the histograms.
  The root mean squared (rms) error of the reconstruction after
  thresholding is indicated as a percentage of the signal rms.  Tick
  marks on the color bars identify the $5^\mathrm{th}$,
  $50^\mathrm{th}$ and $95^\mathrm{th}$ percentile of the coefficients
  or the spatial reconstructions after thresholding, respectively.
  Interior ticks on the histograms roughly coincide with these same
  percentiles as applied to either the positive and negative
  coefficients when expressed on a logarithmic scale.  Histograms for
  the positive coefficients point up and have ordinates in positive
  percentages, histograms for the negative coefficients point down and
  have ordinates in negative percentages; these percentages are with
  respect to the total number of positive and negative coefficients.
  The blue and red shaded areas of the histograms reflect the
  coefficients retained at the global $85^\mathrm{th}$ thresholding
  level.}
\end{figure*}

In Figure~\ref{wavelettopo} we explore coefficient statistics and the
effects of thresholding on the reconstruction errors for a model of
terrestrial topography. We focus on the fifth, or ``North-American''
chunk of our cubed sphere, and use the orthogonal D2 (Haar), D4 and D6
wavelet bases\cite{Daubechies88b} on the interval, with
preconditioning\cite{Cohen+93}. The top row uses the common
conventions in plotting the wavelet and scaling coefficients in each
of the bases after (hard) thresholding\cite{Mallat2008} such that
only the coefficients larger than their value at the 85$^\mathrm{th}$
percentile level survive. The coefficients that have now effectively
been zeroed out are left white in these top three panels. The middle
series of panels of Figure~\ref{wavelettopo} plots the spatial
reconstruction after thresholding at this level; the root mean squared
(rms) errors of these reconstructions are quoted as a percentage of
the original root mean squared signal strengths. The thresholded
wavelet transforms allow us to discard, as in these examples, 85\% of
the numbers required to make a map of North American topography in the
cubed-sphere pixel basis: the percentage error committed is only
6.3\%, 5.2\% and 9.2\% according to this energy criterion in the D2,
D4 and D6 bases, respectively. From the map views it is clear that
despite the relatively small error, the D2 basis leads to unsightly
block artifacts in the reconstruction, which are largely avoided in
the smoother and more oscillatory D4 and D6 bases. A view of the
coefficient statistics is presented in the lowermost three panels of
Figure~\ref{wavelettopo}. The coefficients are roughly log-normally
distributed, which helps explain the success of the thresholded
reconstruction approach. We conclude that the D4 basis is a good
candidate for geophysical data representation.

\begin{figure*}\centering
\includegraphics[width=0.485\columnwidth]{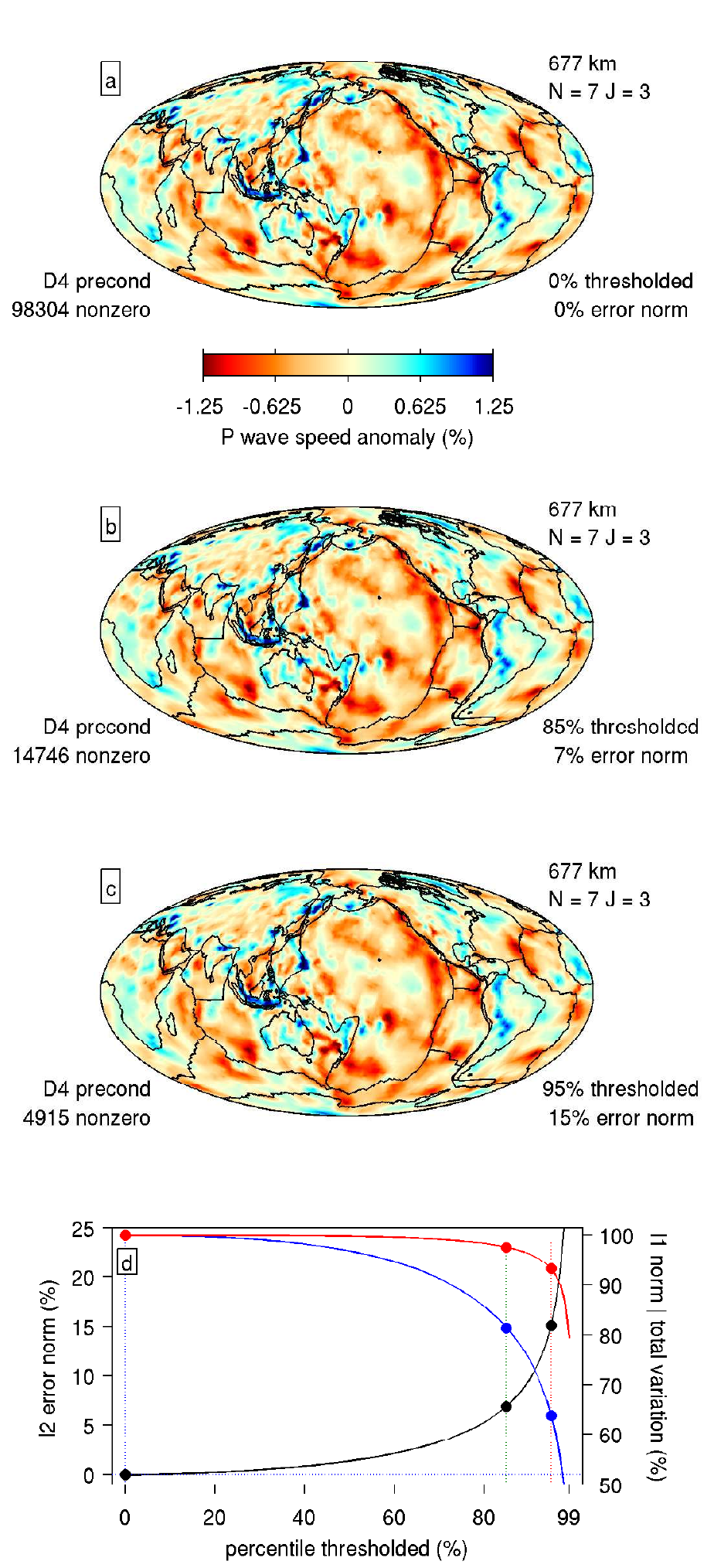}
\includegraphics[width=0.485\columnwidth]{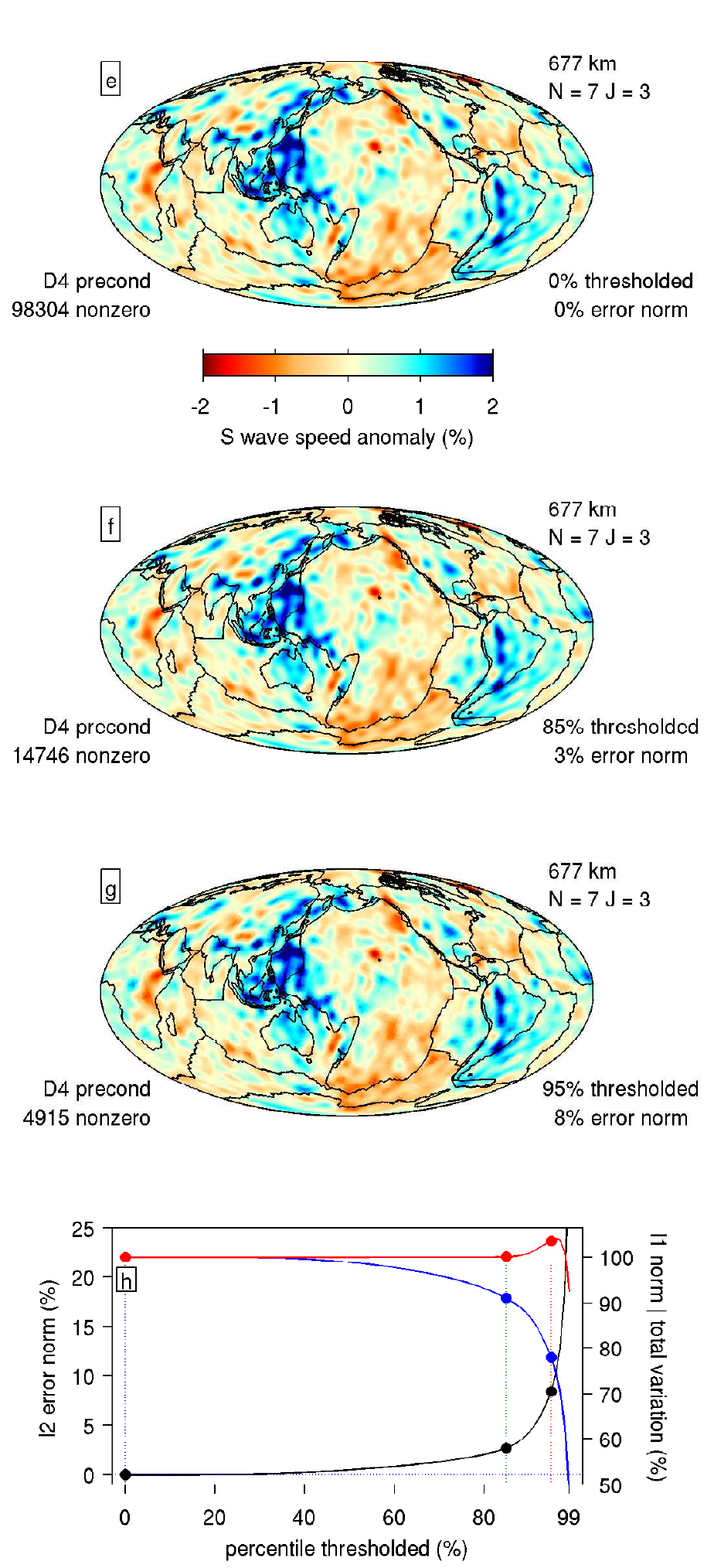}
\caption{\label{waveletmontelli}\label{waveletritsema} Sparsity and
  reconstruction stability of two global seismic wavespeed models
  under incremental hard thresholding of their wavelet and scaling
  coefficients using the preconditioned edge-cognizant D4 wavelet
  basis\cite{Daubechies88b,Cohen+93} in the angular coordinates of the
  cubed sphere, as developed in this paper.  \textit{(a--d)} Results
  for the compressional wave speed seismic model of
  ref.~\citenum{Montelli+2006} and \textit{(e--h)} for the shear wave
  speed seismic model of ref.~\citenum{Ritsema+2010}, at the same
  depth of 677~km below the surface of the Earth, for cubed spheres
  with $6\times 2^{2N}$ elements ($N=7$), and to a $2^J$ dyadic
  subdivision ($J=3$).  As a function of the percentage of the
  coefficients that are being thresholded, and relatively to the
  original unthresholded values, the bottom panels quote the spatial
  $\ell_2$~norms of the reconstruction error (in black), the total
  variation norms of the reconstructed images in the space domain (in
  red), and the $\ell_1$~norms of the coefficients that remain (in
  blue). The reconstructions remain
  faithful to the originals even at elevated levels of thresholding.}
\end{figure*}

For applications in seismology we now illustrate the performance of
the cubed-sphere wavelet basis in compressing seismological Earth
models such as the one by Montelli et al. (2006) of compressional
(\textit{P}) wavespeed heterogeneity and another by Ritsema et
al. (2010) of shear (\textit{S}) wavespeed perturbations. At a depth
of about 670~km, Figs~\ref{waveletmontelli}a and~\ref{waveletritsema}e show
the wavespeed anomalies from the average at that depth. The wavelet
transform in the D4 basis (with special boundary filters and after
preconditioning, and up until scale $J=3$) was thresholded and the
results re-expanded to the spatial grid, identically as we did for the
topography in Figure~\ref{wavelettopo}. The results for specific values
of the thresholding (quoted as the percentile of the original wavelet
coefficients) are shown in Figs~\ref{waveletmontelli}b
and~\ref{waveletritsema}f for the 85$^\mathrm{th}$, and 
Figs~\ref{waveletmontelli}c and~\ref{waveletritsema}g for the
95$^\mathrm{th}$, respectively. At each level of thresholding the
number of nonzero wavelet/scaling expansion coefficients is quoted: at
0\% thresholding this number is identical to the number of pixels in
the surficial cubed sphere being plotted. 

The reconstruction error can be visually assessed from the pictures;
it is also quoted next to each panel as the percentage of the root
mean squared error between the original and the reconstruction,
normalized by the root mean squared value of the original in the
original pixel representation, in percent. Specifically, we calculate
and quote the ratio of $\ell_2$~norms in the pixel-basis model vector~$\bm$,
\be\label{l2error}
 100\times \left.\left\|\bm-\mS%\!
     \left\{\mT\!\left[\mW(\bm)\right]
     \right\}\right\|_2\right/ \left\|\bm\right\|_2, 
\ee 
which, in the lower-right annotations is called the ``\% error norm''.
We write $\mW$ for any of the wavelet (analysis) transforms that can
be used and $\mS$ (synthesis) for their inverses, and $\mT$ for the
hard thresholding of the wavelet and scaling
coefficients. In Figs~\ref{waveletmontelli}e
and~\ref{waveletmontelli}j, the misfit quantity~(\ref{l2error})
is represented as a black line relevant to the left ordinate labeled
``$\ell_2$~error norm'', which shows its behavior at 1\% intervals of
thresholding; the filled black circles correspond to the special cases
shown in the map view. Only after about 80\% of the coefficients have
been thresholded does the error rise above single-digit percentage
levels, but after that, the degradation is swift and inexorable. The
blue curves in Figs~\ref{waveletmontelli}e and~\ref{waveletmontelli}j
show another measure relevant in this context, namely the ratio of the
$\ell_1$~norms of the thresholded wavelet coefficients compared to the
original ones, in percent, or \be\label{l1error} 100\times
\left.\left\| \mT\!\left[\mW(\bm)\right] \right\|_1\right/
\left\|\mW(\bm)\right\|_1.  \ee The $\ell_2$ ratios~(\ref{l2error}) in
the black curves (and the left ordinate) evolve roughly symmetrically
to the $\ell_1$ ratios~(\ref{l1error}) in the blue curves (and the
right ordinate), though evidently their range is different. The third
measure that is being plotted as the red curve is the ``total
variation'' norm ratio, in percent, namely 
\be\label{tverror}
100\times \left.\left\|
\bnabla\mS%\!
    \left\{\mT\!\left[\mW(\bm)\right]
    \right\}
\right\|_1\right/ 
\left\| 
\bnabla\bm
\right\|_1,
\ee
whereby $\left\|\bnabla\bm\right\|_1$ is the sum over all voxels of
the length of the local gradient of~$\bm$. By this
measure, which is popular in image restoration
applications\cite{Rudin+92,Dobson+96,Chambolle+97}, the quality of
the reconstruction stays very high even at very elevated levels of
thresholding; we note that its behavior is not monotonic and may
exceed~100\%. 

\begin{figure*}\centering
\includegraphics[width=0.825\textwidth]{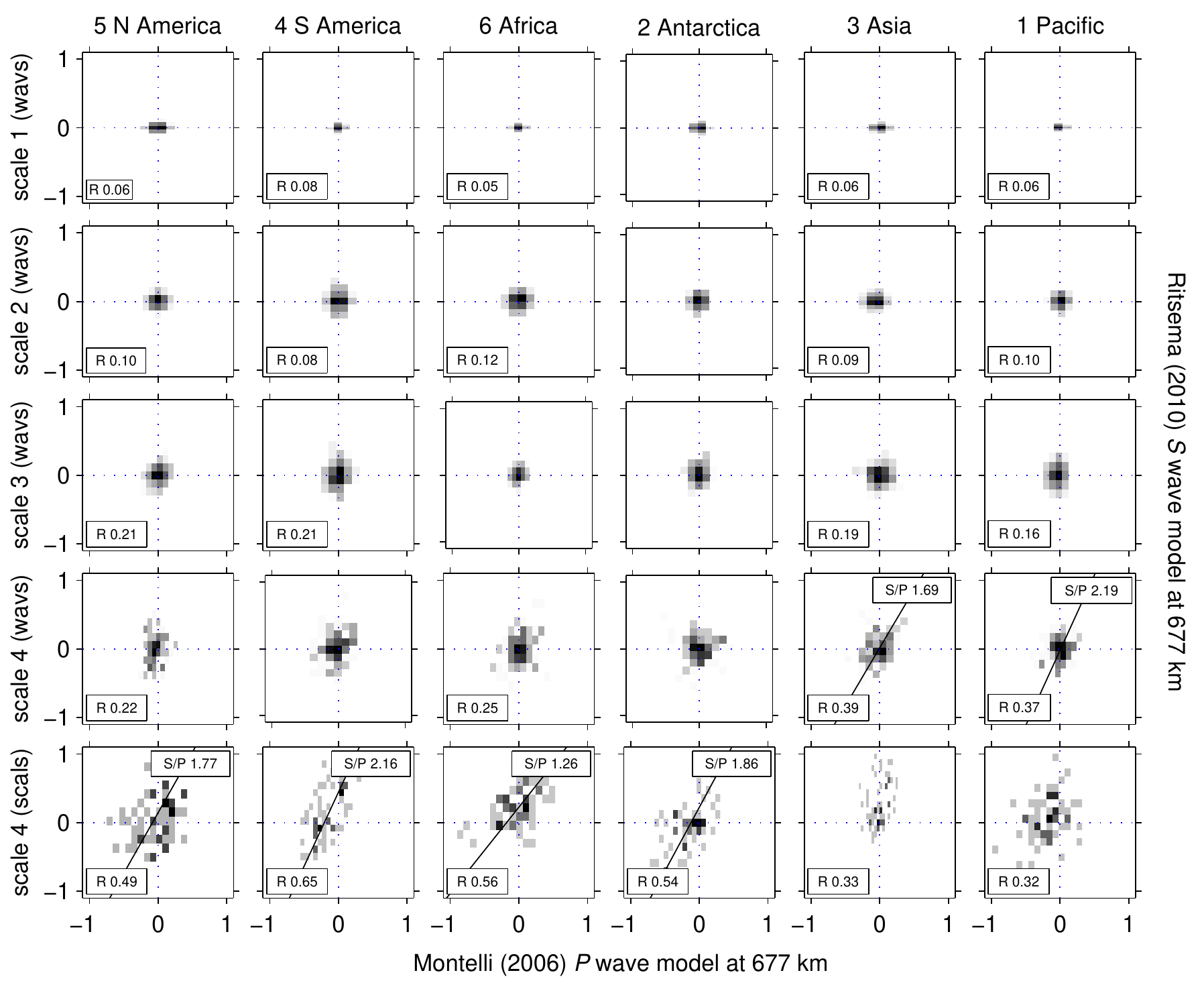} 
\caption{\label{psplots}Joint properties of seismic mantle structure
  in the Montelli (2006) \textit{P}-wave and Ritsema (2010)
  \textit{S}-wave speed models, at 677~km depth in the Earth. Every
  row corresponds to a different scale in the D4 wavelet decomposition
  of the models. Each panel shows the logarithmic density of
  observations. Black shading corresponds to the maximum 
  density in each panel; all patches that account for less than 1\% of
  the observations are rendered white. Using total least squares a
  regression line was fit to all sets with correlation coefficients
  exceeding 0.35. The slope of the line, a measure of the $\delta\ln
  V_S/\delta\ln V_P$ ratio appears in the top right
  corners. Significant correlation coefficients are quoted in the bottom left.}
\end{figure*}

Finally, the new wavelet construction can be used to study the
joint properties of the wavelet coefficients of seismic wavespeed
models, as we illustrate in Figure~\ref{psplots}. There, we report the
correlation between wavelet coefficients in the Montelli and Ritsema
models as a function of scale and approximate geographical position
(see again Figure~\ref{twodtopo} for the numbering scheme of the
cubed-sphere chunk). A rendering of the two-dimensional density of the
data is accompanied by the value of their correlation coefficient
(lower left labels) where this is deemed significant at the 95\%
level, and the slope of the total-least-squares based fit in this
space (upper right labels), which is only quoted when the correlation
coefficients exceeded 0.35. This should provide an estimate of the
logarithmic ratio of shear-wave to compressional-wave speed
perturbations, $\delta\ln V_S/\delta\ln V_P$, an important
discriminant in the interpretation of the (thermal or chemical) cause
of seismic velocity anomalies \cite{Trampert+2005}. The 
variation of this ratio as a function of scale and chunk position
yields information that will be of use for geochemical and
geodynamical studies, and the orthogonality of the wavelet basis in
scale and physical space removes some of the arbitrariness in the
calculation. Ritsema's model does not have much structure at the
smallest scales. From scale~3 onward a positively correlated pattern
begins to emerge, though even at this particular scale, the
correlation coefficients remain below the relatively stringent 0.35
level. Wavelets and scaling coefficients are well correlated at
the largest scale~4 considered, with several of the correlation coefficients
exceeding our threshold. This information is useful to
geophysicists\cite{Tkalcic+2002,Saltzer+2001,Deschamps+2003,Trampert+2005},
especially given our new-found ability to study the regional variation
of such ratios, taking into account their dependence on scale length. 

Simons et al. (2011) detail the reasoning behind the construction of
the ``superchunk'' cubed sphere shown in  Figure~\ref{thechunk}
(\textit{right}). In a nutshell, the cubed-sphere bases
are to be used not simply for the representation and analysis of
seismic models, but also to parameterize the inversion of primary data
for such models.  Since the edge-cognizant transforms are not
norm-preserving, the thresholding steps involved in using common
algorithms such as
(F)ISTA\cite{Loris+2007,Beck+2008,Loris+2010}, if
unmodified, lead to artifacts in the solution which are largely
avoided by building the wavelet transforms in the superchunk domain
and ignoring the edges altogether\cite{Simons+2011b}.

\section{The Spherical Slepian Tree Transform}
\label{seccon}

Geophysical and cosmological signals are constrained by the physical
processes that generate them and conditioned by the sensors that
observe them. On the real line and in the plane, both physical and
sampling constraints lead to assumptions of a bandlimit: that a signal
contains no energy outside some supporting region in the frequency
domain. Bandlimited signals on the sphere are zero save for the
low-frequency spherical-harmonic components.  Spherical harmonics are
not orthogonal on arbitrary portions of the sphere, yet for the study
of geophysical processes we may not have access to nor interest in
signal originating from outside a specific geographic region of
interest. By construction, Slepian functions satisfy the dual
constraints of bandlimitation and spatial concentration by
optimization of an energy criterion that concentrates as much energy
as possible into the region of interest for a given
bandlimit\cite{Simons+2006a}. With Slepian function bases, however, it
is ``all or nothing'': given a certain bandwidth and a spatial region
of interest, the functions (variably) fill the entire bandwidth range
and ultimately the entire spatial target. The wavelet constructions
that we introduced in Section~\ref{firstcon}, on the other hand, were
compactly supported and had multiresolution properties which the
Slepian basis does not possess.

Here we develop an algorithm for the construction of Slepianesque
dictionary elements that are bandlimited, localized, and multiscale.
It is based on a subdivision scheme that constructs a
binary tree from subdivisions of the region of interest. Such a
dictionary has many nice properties: it closely overlaps with the most
concentrated Slepian functions on the region of interest, and most
element pairs have low coherence. Therefore, they, too, should be
eminently suitable to solve ill-posed inverse problems in geophysics
and cosmology. Though the new dictionary is no longer composed of
purely orthogonal elements like the Slepian basis, it can also be
combined with modern inversion techniques that promote sparsity in the
solution. Thereby they provide significantly lower residual error
after reconstruction when compared to ``classically optimal'' Slepian
inversion techniques\cite{Simons+2006b}.

\subsection{A Multiscale Dictionary  of  Slepian Functions}

\begin{figure}[h]
\centering
\includegraphics[width=.5\linewidth]{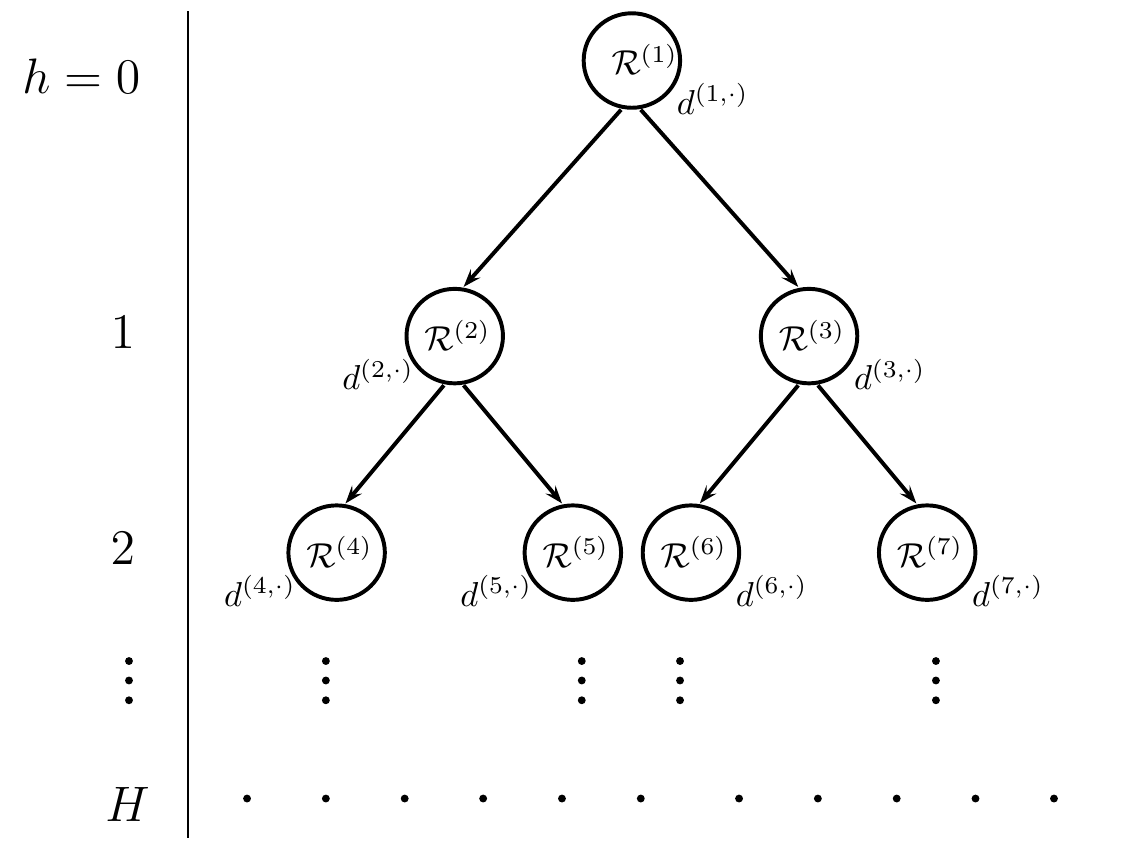}
\caption[The binary tree subdivision scheme and associated
  dictionary~$\cD_{\cR,L}$.]{\label{fig:sltrdiag}The binary tree
  subdivision scheme and associated dictionary~$\cD_{\cR,L}$ for the
  multiscale Slepian tree transform.  We
  define the top-level region $\cR$ as $\cR^{(1)}$ and the generic
  subsets $\cR'$ as $R^{(j)}$.}
\end{figure}

\begin{figure}[h]
\centering
\begin{subfigure}[$d^{(1,1)}$]{
\includegraphics[width=.275\textwidth]{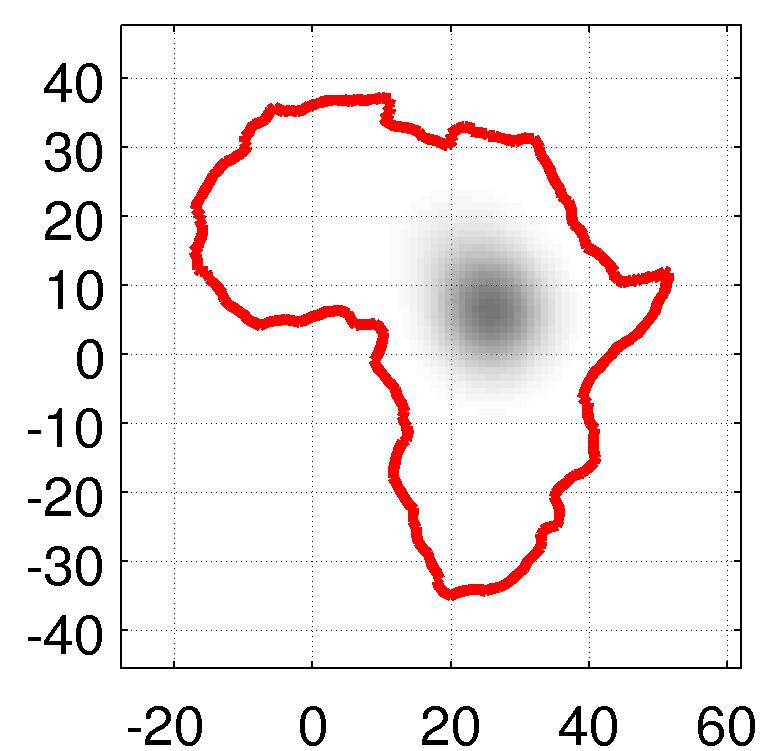}}
\end{subfigure}
\begin{subfigure}[$d^{(2,1)}$]{
\includegraphics[width=0.27\textwidth]{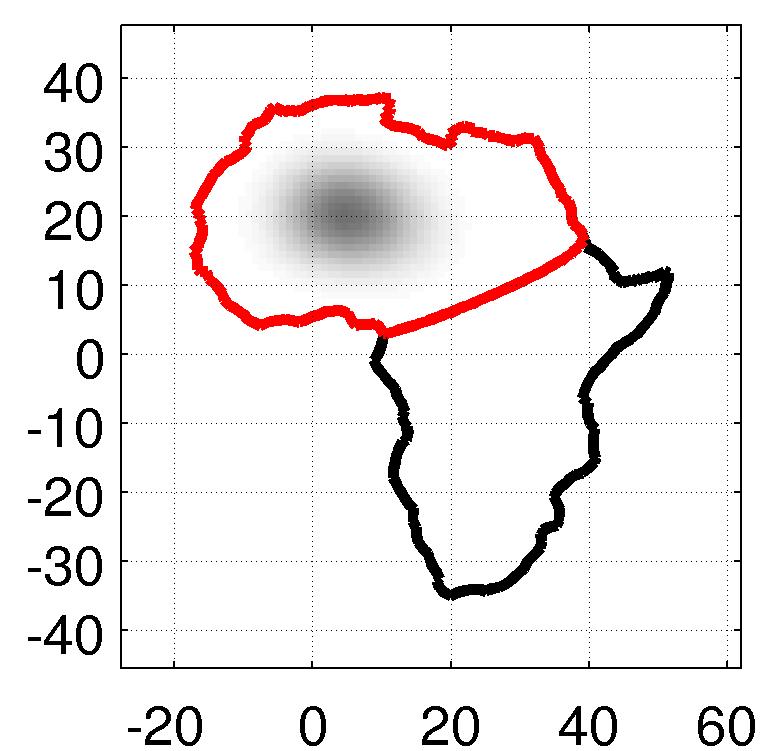}}
\end{subfigure}
\begin{subfigure}[$d^{(3,1)}$]{
\includegraphics[width=0.27\textwidth]{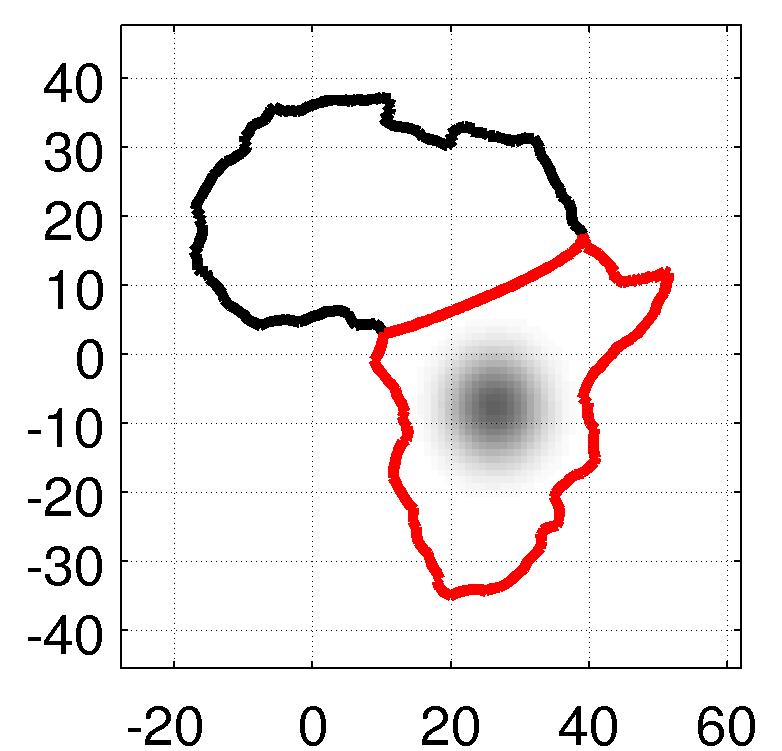}}
\end{subfigure}
\\
\begin{subfigure}[$d^{(4,1)}$]{
\includegraphics[width=0.27\textwidth]{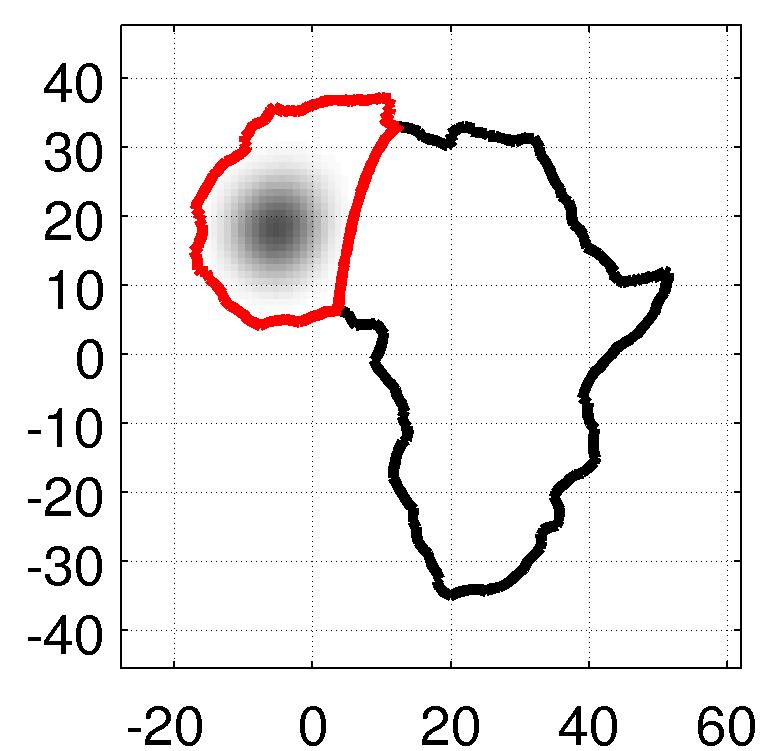}}
\end{subfigure}
\begin{subfigure}[$d^{(5,1)}$]{
\includegraphics[width=0.27\textwidth]{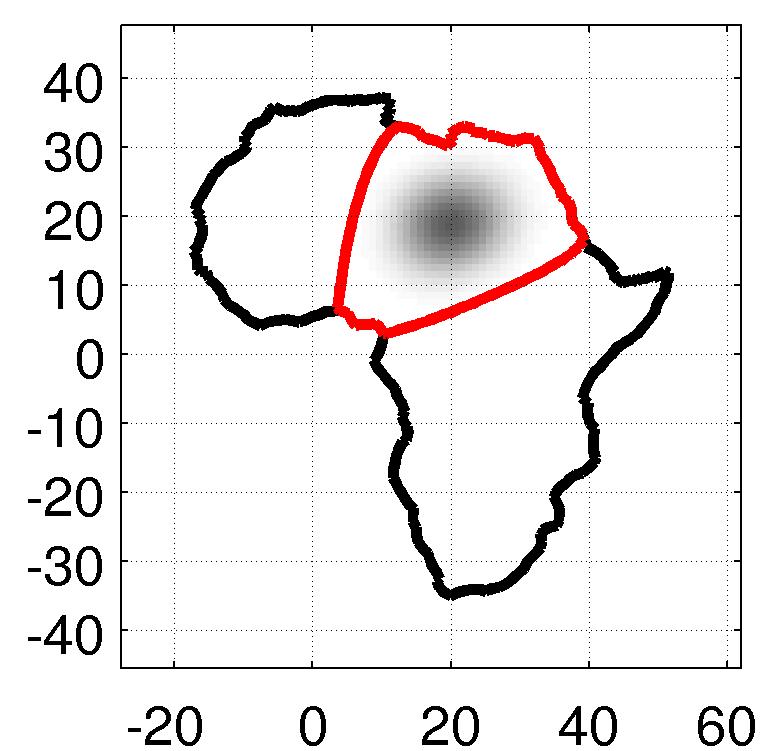}}
\end{subfigure}
\begin{subfigure}[$d^{(6,1)}$]{
\includegraphics[width=0.27\textwidth]{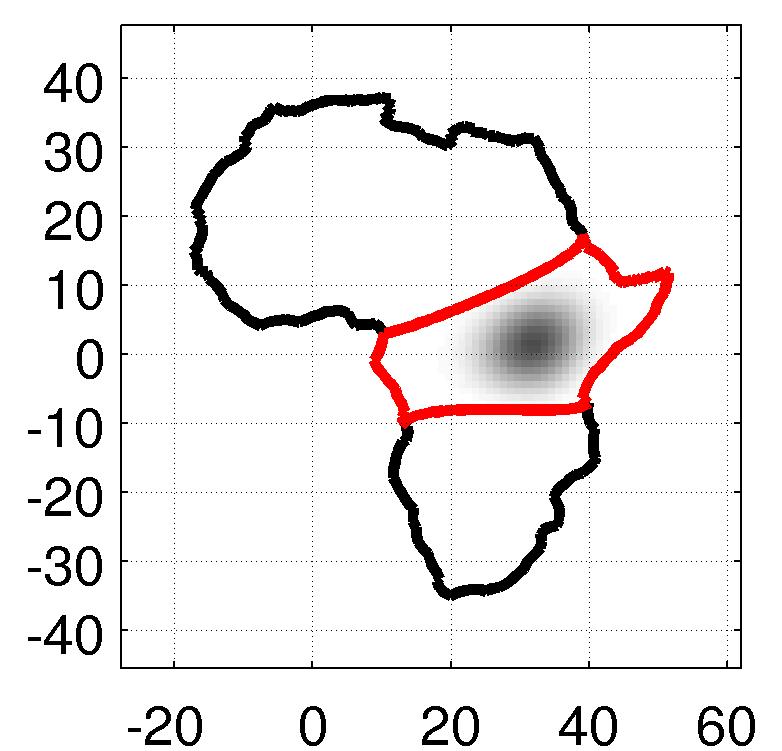}}
\end{subfigure}
\begin{subfigure}[$d^{(250,1)}$]{
\includegraphics[width=0.27\textwidth]{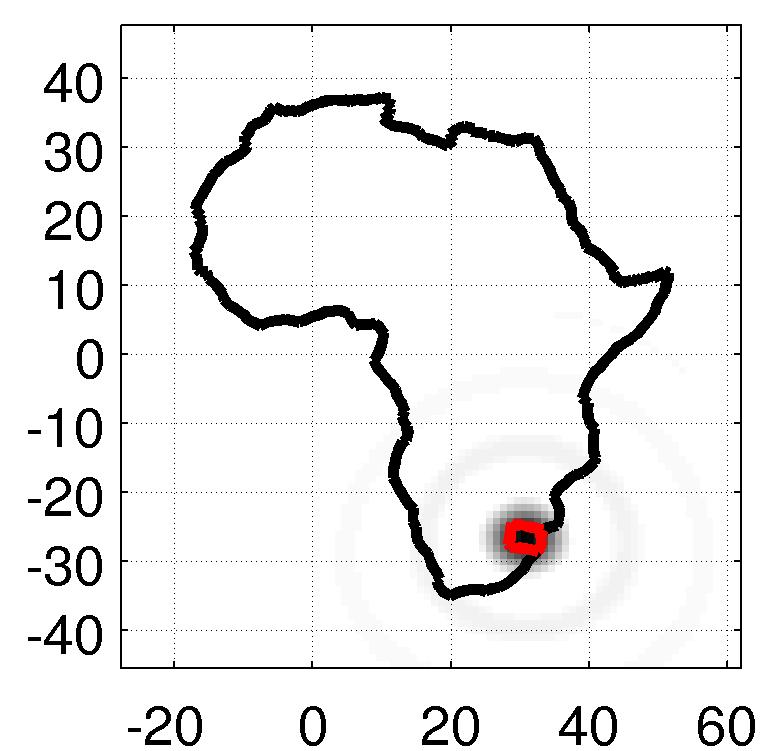}}
\end{subfigure}
\begin{subfigure}[$d^{(251,1)}$]{
\includegraphics[width=0.27\textwidth]{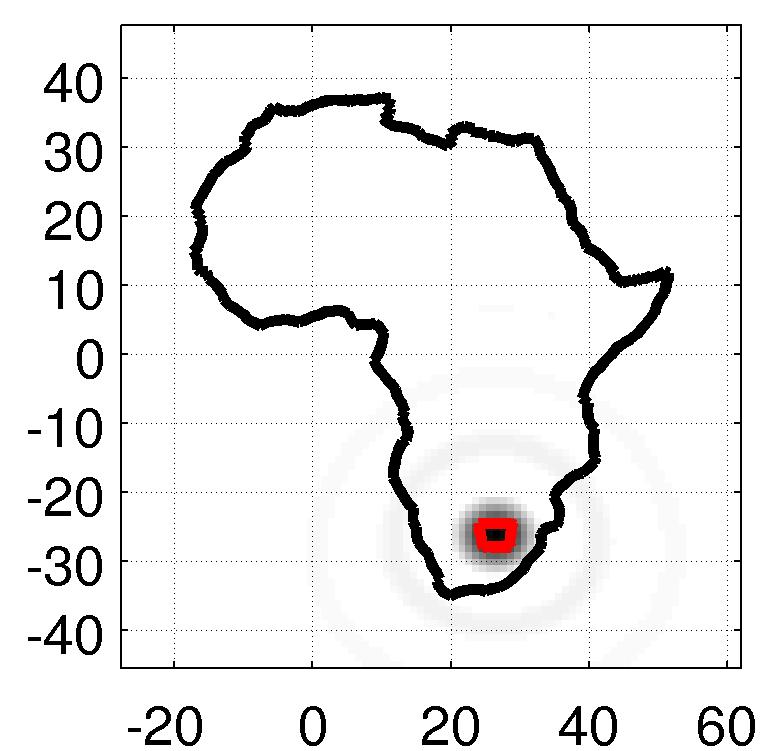}}
\end{subfigure}
\begin{subfigure}[$d^{(252,1)}$]{
\includegraphics[width=0.27\textwidth]{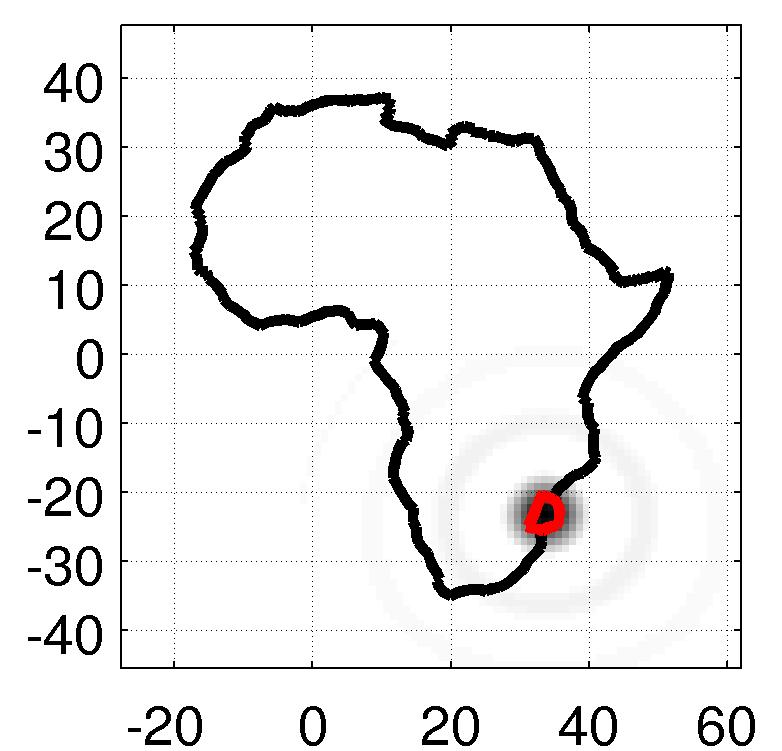}}
\end{subfigure}
\\
\begin{subfigure}[$d^{(253,1)}$]{\includegraphics[width=0.27\textwidth]{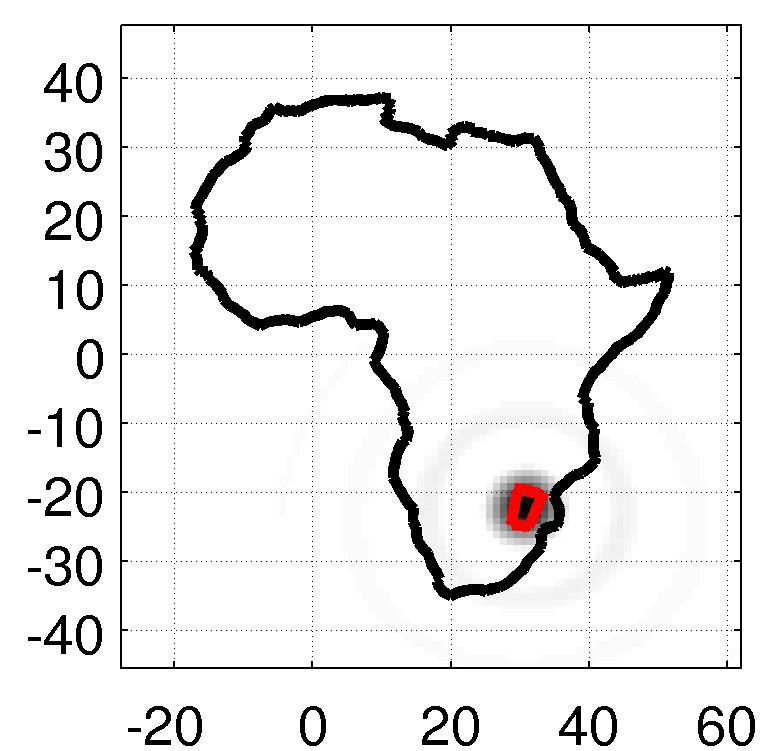}}
\end{subfigure}
\begin{subfigure}[$d^{(254,1)}$]{
\includegraphics[width=0.27\textwidth]{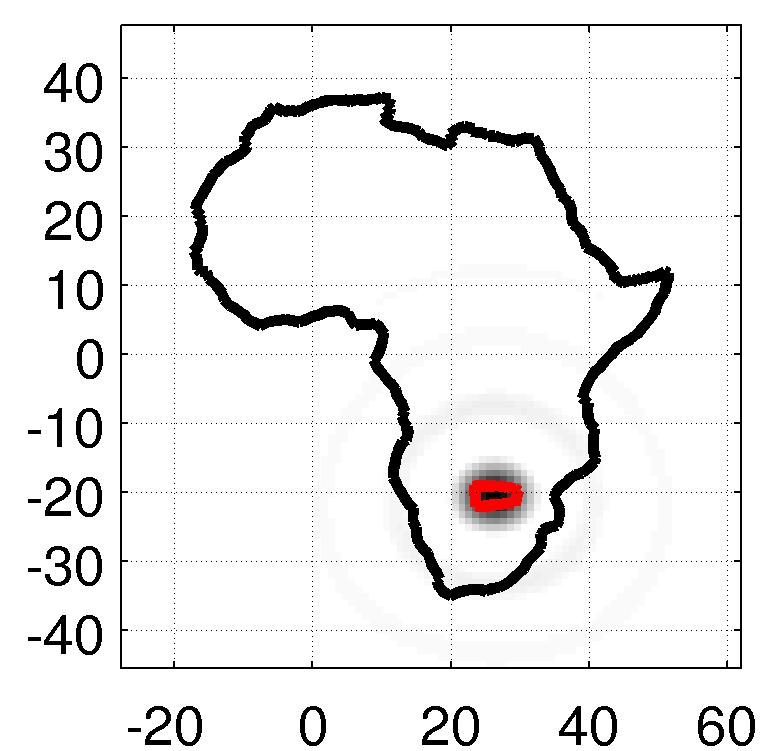}}
\end{subfigure}
\begin{subfigure}[$d^{(255,1)}$]{
\includegraphics[width=0.27\textwidth]{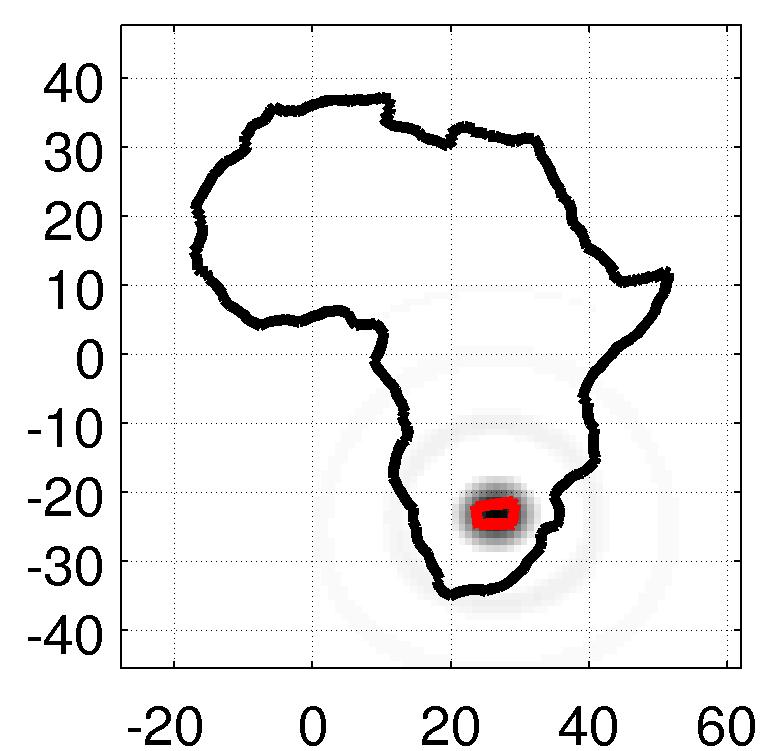}}
\end{subfigure}
\caption
{\label{fig:sleptrafrica}Slepian tree dictionary
  $\cD_{\textrm{\small Africa},36,1}$ of size $\abs{\cD}=255$ showing the 
  functions $d^{(1,1)}$ through $d^{(6,1)}$ and $d^{(250,1)}$ through
  $d^{(255,1)}$.
  The x-axis is longitude, the y-axis is colatitude.  Regions of
  concentration $\set{\cR^{(i)}}$ are outlined.}
\end{figure}

We now turn our focus to numerically constructing a dictionary $\cD$
of functions that can be used to approximate bandlimited signals on
the sphere and allows for the reconstruction of signals from their
point samples. Let $\cR \subset S^2$ be a simply connected subset of
the sphere.  With~$L$ the bandwidth, the dictionary $\cD$ will be
composed of functions bandlimited to spherical harmonic degrees $0
\leq l \leq L$. The construction is based on a binary tree. The node
capacity, $n$, is a positive integer. Each node of the tree
corresponds to the first $n$ Slepian functions with bandlimit $L$ and
concentrated on a subset $\cR' \subset \cR$.  The top tree node is for
the entire region $\cR$, and each node's children correspond to a
division of $\cR'$ into two roughly equally sized subregions.  As the
child nodes will be concentrated in disjoint subsets of $\cR'$, all of
their corresponding functions and children are effectively incoherent.
We now fix a height $H$ of the tree: the number of times to subdivide
$\cR$.  The height is determined as the maximum number of binary
subdivisions of $\cR$ that can have $n$ well concentrated functions.
That is, we find the minimum integer $H$ such that $n \geq N_{2^{-H}
  \abs{\cR}, L}$, with $N$ the Shannon number\cite{Simons+2006a},
which has the solution
$$
H = \ceil{\log_2\left(\frac{\abs{\cR}}{4 \pi} \frac{(L+1)^2}{n}\right)}.
$$
A complete binary tree with height $H$ has $2^{H+1}-1$ nodes, so from
now on we will denote the dictionary 
$$
\cD_{\cR,L,n} =
\set{d^{(1,1)},d^{(1,2)},\cdots,d^{(1,n)},\cdots,d^{\left(2^{H+1}-1,1\right)},
  \cdots,d^{\left(2^{H+1}-1,n\right)}}
$$
as the set of
$\abs{\cD_{\cR,L,n}} = n \, (2^{H+1}-1)$ functions thus constructed
on region $\cR$ with bandlimit $L$ and node capacity
$n$.
Figure~\ref{fig:sltrdiag} shows the tree diagram of the
subdivision scheme.  We use the standard enumeration of nodes wherein
node $(j,\cdot)$ is subdivided into child nodes $(2j,\cdot)$ and
$(2j+1,\cdot)$, and at a level $0 \leq h \leq H$, the nodes are
indexed from $2^h \leq j \leq 2^{h+1}-1$.  More specifically, for
${j = 1,2,\ldots}$, we have ${\cR^{(j)} = \cR^{(2j)} \cup \cR^{(2j+1)}}$.
Furthermore, letting $g^{\cR'}_\alpha$ be the $\alpha^\mathrm{th}$ Slepian
function on $\cR'$ (the solution to the classical Slepian
concentration criterion eq.~4.1 in ref.~\citenum{Simons+2006a}, with
concentration region $\cR'$), we have that
$$
d^{(j,\alpha)} = g^{\cR^{(j)}}_\alpha.
$$
Figure~\ref{fig:sleptrafrica} shows an example of the construction when
$\cR$ is the African continent.  Note how, for example, $d^{(4,1)}$
and $d^{(5,1)}$ are the first Slepian functions associated with
the subdivided domains of $\cR^{(2)}$. 

\begin{figure}[h]
\centering
\begin{subfigure}[$d^{(1,2)}$]{
\includegraphics[width=0.2775\linewidth]{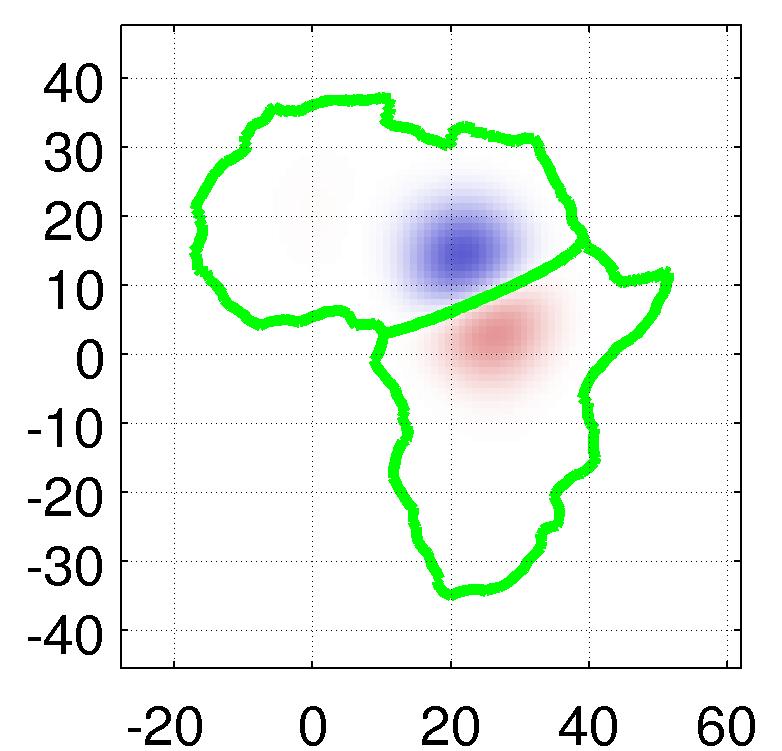}}
\end{subfigure}
\begin{subfigure}[$d^{(2,2)}$]{
\includegraphics[width=0.2775\linewidth]{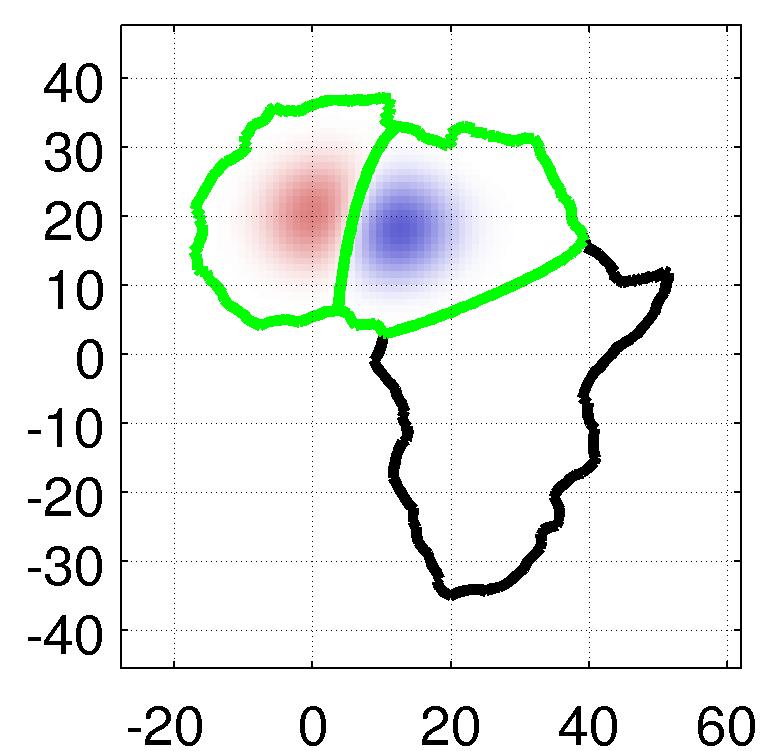}}
\end{subfigure}
\begin{subfigure}[$d^{(3,2)}$]{
\includegraphics[width=0.2775\linewidth]{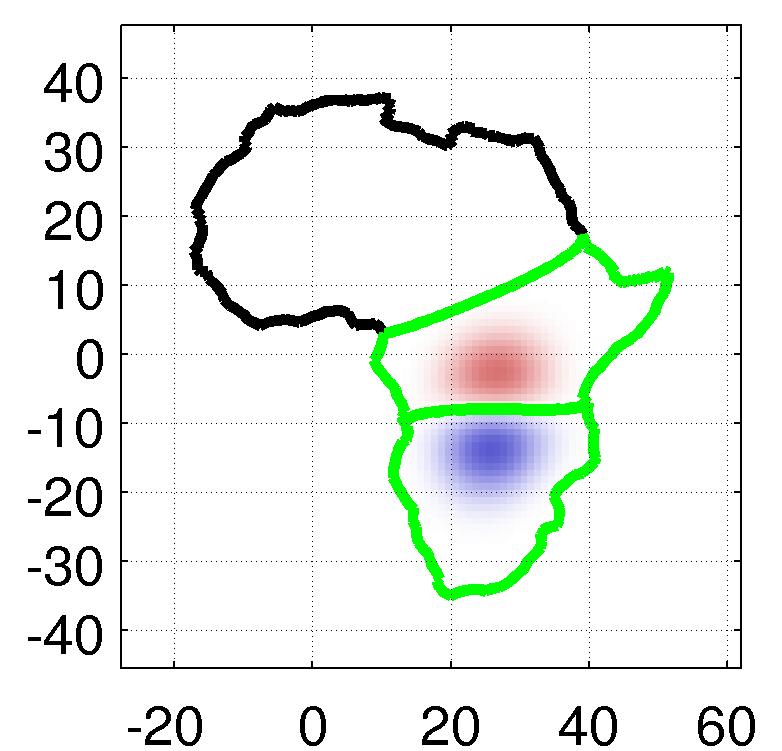}}
\end{subfigure}
\\
\begin{subfigure}[$d^{(4,2)}$]{
\includegraphics[width=0.2775\linewidth]{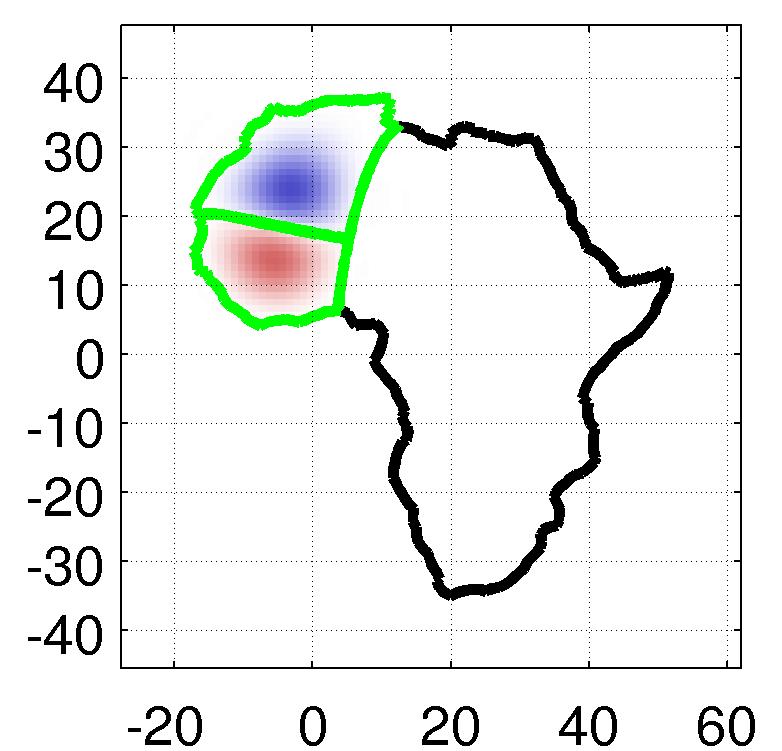}}
\end{subfigure}
\begin{subfigure}[$d^{(5,2)}$]{
\includegraphics[width=0.2775\linewidth]{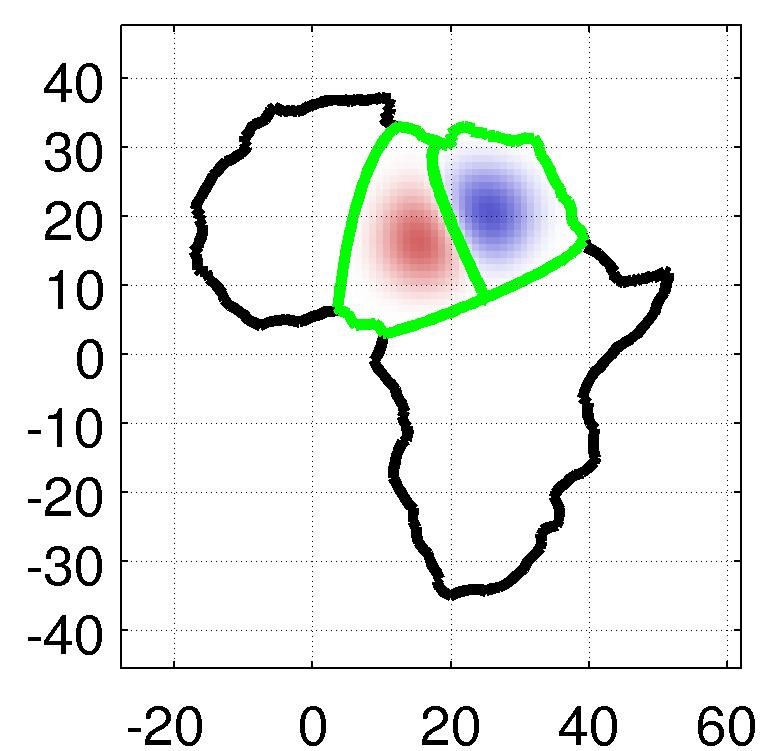}}
\end{subfigure}
\begin{subfigure}[$d^{(6,2)}$]{
\includegraphics[width=0.2775\linewidth]{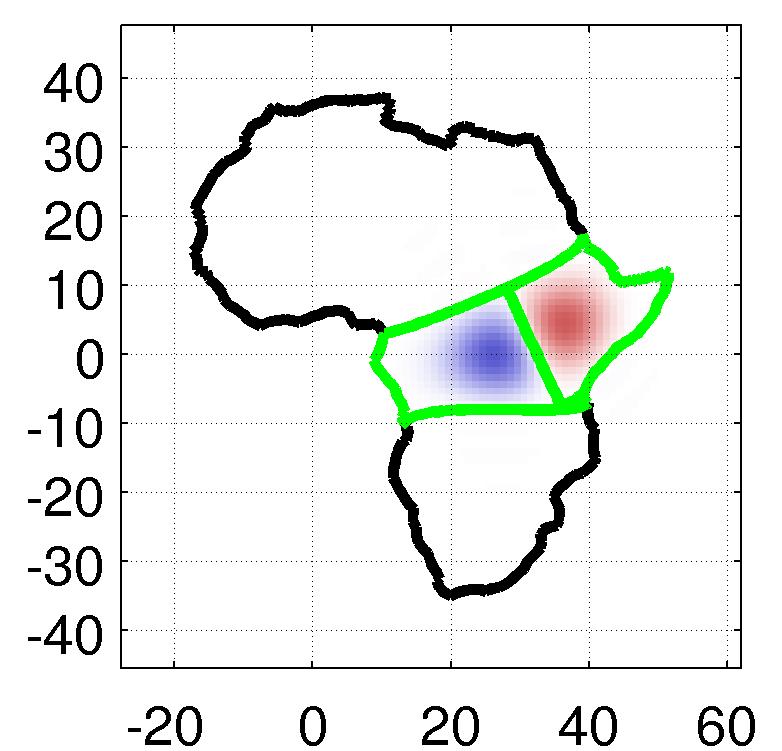}}
\end{subfigure}
\caption
{\label{fig:sleptrafricasecond}Second ($\alpha=2$) Slepian functions associated
  with the regions in Figure~\ref{fig:sleptrafrica}a--f. Regions of
  concentration, and the dividing contour (the zero-level set),
  are drawn in green. Blue and red represent the sign of the
  Slepian function values.}
\end{figure}

To complete the top-down construction, it remains to decide how to
subdivide a region $\cR'$ into equally sized subregions.  For roughly
circular connected domains, the first Slepian function has no sign
changes, and the second Slepian function has a single zero-level curve
that subdivides the region into approximately equal areas; when $\cR'$
is a spherical cap, the subdivision is exact~\cite{Simons+2006a}.  We
thus subdivide a region $\cR'$ into the two nodal domains associated
with the second Slepian function on that domain; see
Figure~\ref{fig:sleptrafricasecond} for a visualization of the
subdivision scheme as applied to the African continent.

\subsection{Concentration, Range, and Incoherence}

The utility of the tree construction presented above depends on its
ability to represent bandlimited functions in a region $\cR$, and its
efficacy at reconstructing functions from point samples in $\cR$.
These properties, in turn, reduce to questions of concentration,
range, and incoherence. First, dictionary $\cD$ is concentrated in $\cR$
if its functions are concentrated in $\cR$. Second, the range of
dictionary $\cD$ is the subspace spanned by its elements. Ideally, the
basis formed by the first few Slepian functions on $\cR$ is a subspace
of the range~of~$\cD$. Third, when $\cD$ is incoherent, pairwise inner
products of its elements have low amplitude: pairs of functions are
approximately orthogonal. This, in turn, is a useful property when
using $\cD$ to estimate signals from point samples.

Unlike the concentration eigenvalues corresponding to the Slepian
functions on $\cR$, not all of the eigenvalues of the elements of
$\cD_{\cR}$ reflect their concentration within this top-level (parent)
region.  We thus define the modified concentration value
\be
\nu^{(j,\alpha)} = \int_{\cR} \left[d^{(j,\alpha)}(x)\right]^2 d\mu(x).
\ee
Recalling that $\norm{d^{(j,\alpha)}}_2 = 1$, the value
$\nu$ is simply the percentage of energy of the
$(j,\alpha)^{\textrm{th}}$ element that is concentrated in $\cR$.  This value is
always larger than the element's eigenvalue,
which relates its fractional energy within the smaller subset $\cR^{(j)}$.

The size of dictionary $\cD_{\cR,L,n}$ is generally larger than the
Shannon number $N_{\abs{\cR},L}=(L+1)^2/\abs{\cR}/4/\pi$ for any node
capacity $n$, and as a result it cannot form a proper basis: it has
too many functions.  Ideally, then, we require that elements of the
dictionary span the space of the first $N_{\abs{\cR},L}$ Slepian
functions. One possible answer to the question of the range of $\cD$
is given by studying the angle between the subspaces~\cite{Wedin1983}
spanned by elements of $\cD$ and the first $\alpha$ functions of the
Slepian basis, for ${\alpha=1,2,\ldots}$.  The angle between two
subspaces $A$ and $B$, possibly with different dimensions, is given by
the formula \be \angle(A,B) = \min\left(\sup_{x \in A} \angle(x,B),
  \sup_{y \in B} \angle(y,A)\right),
\quad \mbox{where} \\
\angle(x,B) = \inf_{y \in B} \angle(x,y) = \cos^{-1} \frac{\norm{P_B
    x}}{\norm{x}}.  \ee Here, $P_B$ is the orthogonal projection onto
space $B$ and all of the norms are with respect to the given subspace.
The angle $\angle(A,B)$ is symmetric, non-negative, and zero iff $A
\subset B$ or $B \subset A$; furthermore it is invariant under unitary
transforms applied to both on $A$ and $B$, and admits a triangle
inequality.  It is thus a good indicator of distance between two
subspaces; furthermore, it can be calculated accurately given two
matrices whose columns span $A$ and $B$. We can therefore identify the
matrices $A$ and $B$ with the subspaces spanned by their columns.

\begin{figure}[ht]
\centering
\includegraphics[width=.4725\textwidth]{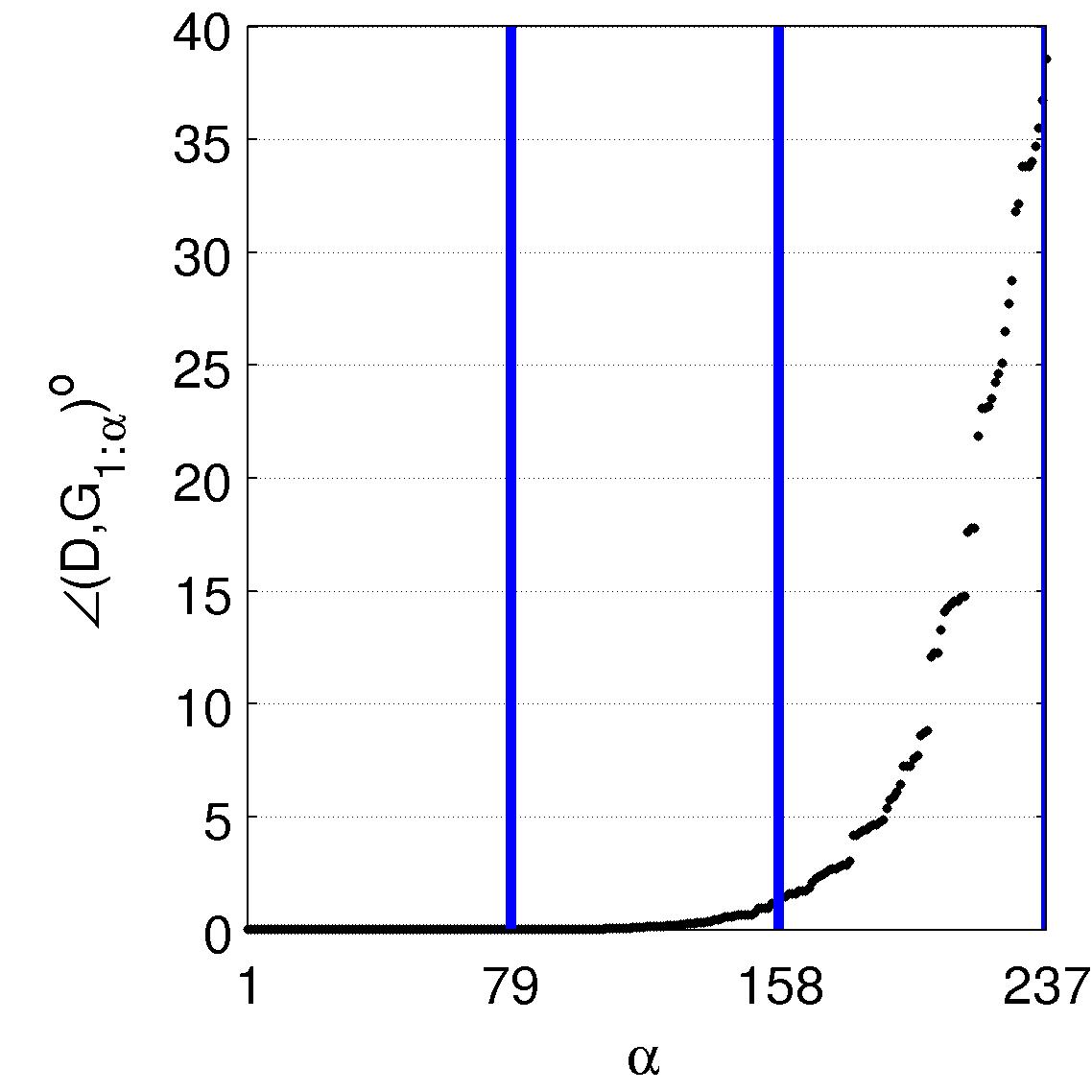}
\includegraphics[width=.4725\textwidth]{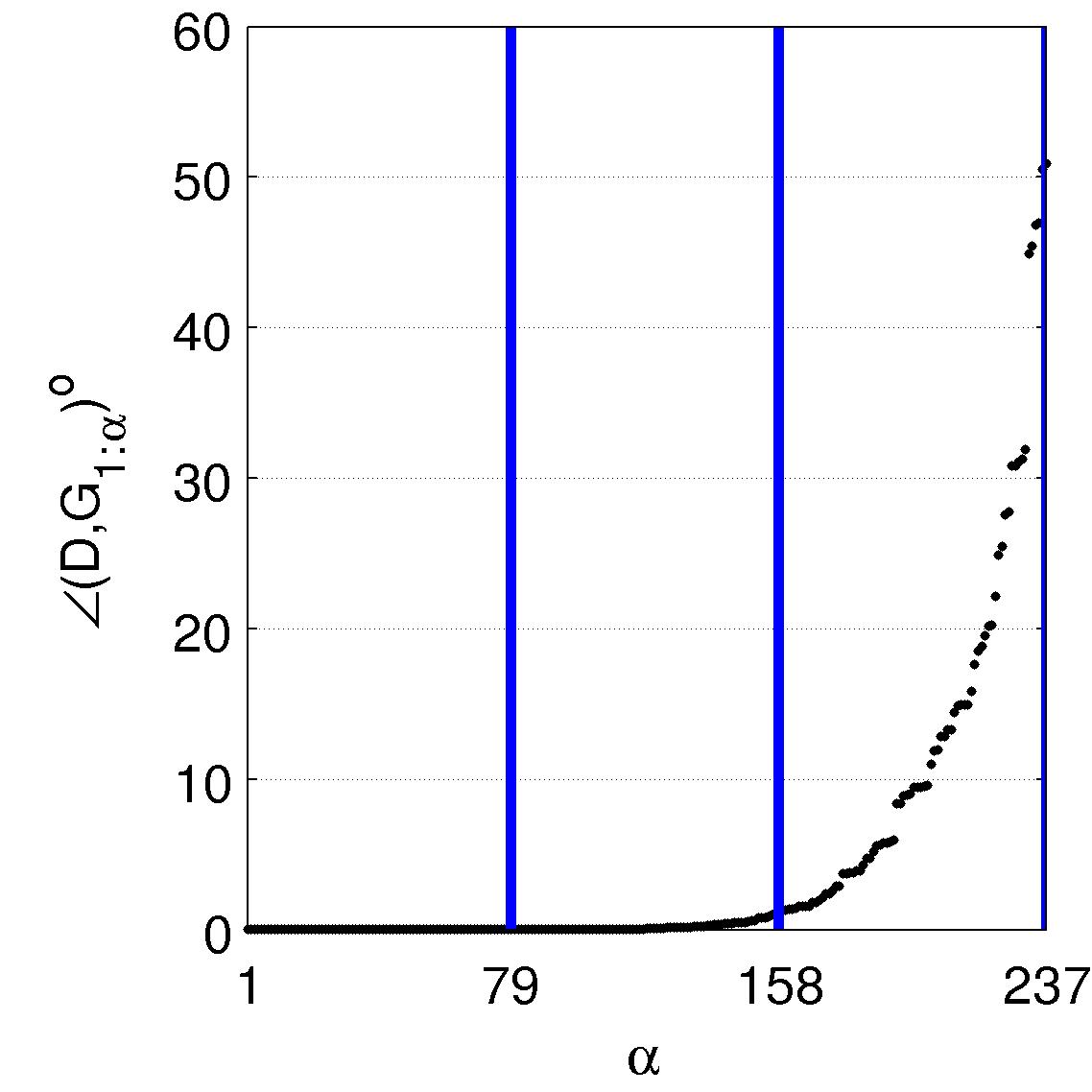}
\caption{\label{fig:slepvstrdist}Angles (in degrees)
  between the spaces spanned by
  $\big(\wh{G}_{\abs{\cR},36}\big)_{1:\alpha}$ and the dictionary
  matrices $\wh{D}_{\cR,36,1}$ (\textit{left}) and $\wh{D}_{\cR,36,2}$ (\textit{right}),
  $\cR=\textrm{\small Africa}$, $\alpha=1,\ldots,3 N_{\abs{\cR},L}$.
  Thick lines correspond to integer multiples of the Shannon number~%
  ${N_{\textrm{\small Africa},36} \approx 79}$.}
\end{figure}

Let $\big(\wh{G}_{\cR,L}\big)_{1:\alpha}$ denote the matrix containing the
first $\alpha$ column vectors which are the spherical harmonic
expansion coefficients of the traditional Slepian functions for a
given region $\cR$ and a bandwidth~$L$.  Let $\wh{D}$ denote the
$(L+1)^2 \x \abs{\cD_{\cR,L,n}}$ matrix 
containing the spherical harmonic representations of the elements of
the dictionary~$\cD_{\cR,L,n}$. Figure~\ref{fig:slepvstrdist} shows
$\angle(\wh{G}_{1:\alpha}, \wh{D})$ for ${\cR=\textrm{\small Africa}}$ with ${L=36}$. 
The Shannon number is ${N_{\textrm{\small Africa},36} \approx 79}$.
It is clear that while the dictionaries
$\cD_{\textrm{\small Africa},36,1}$ and $\cD_{\textrm{\small Africa},36,2}$ do not
strictly span the space of functions bandlimited to $L=36$ and
optimally concentrated in Africa, they are a close approximation: the
column span of $\big(\wh{G}_{\textrm{\small Africa},36}\big)_{1:\alpha}$ is
nearly linearly dependent with the spans of
$\wh{D}_{\textrm{\small Africa},36,1}$ and $\wh{D}_{\textrm{\small Africa},36,2}$, for
$\alpha$ significantly larger than $N_{\textrm{\small Africa},36}$.

\begin{figure}[ht]
\centering
\begin{subfigure}[$\cD_{\textrm{\small Africa},36,1}$]{
\includegraphics[width=0.485\textwidth]{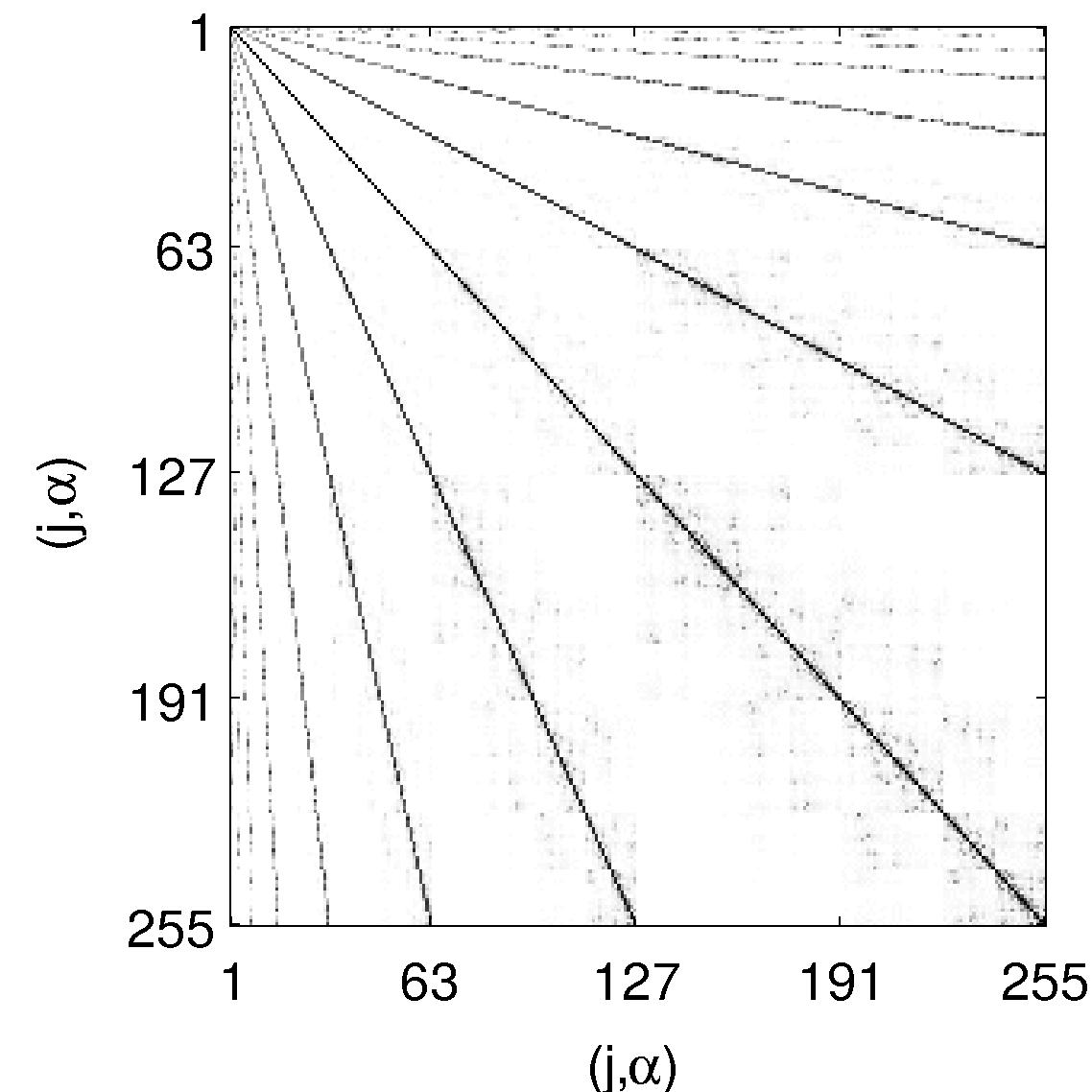}}
\end{subfigure}
\begin{subfigure}[$\cD_{\textrm{\small Africa},36,2}$]{
\includegraphics[width=0.485\textwidth]{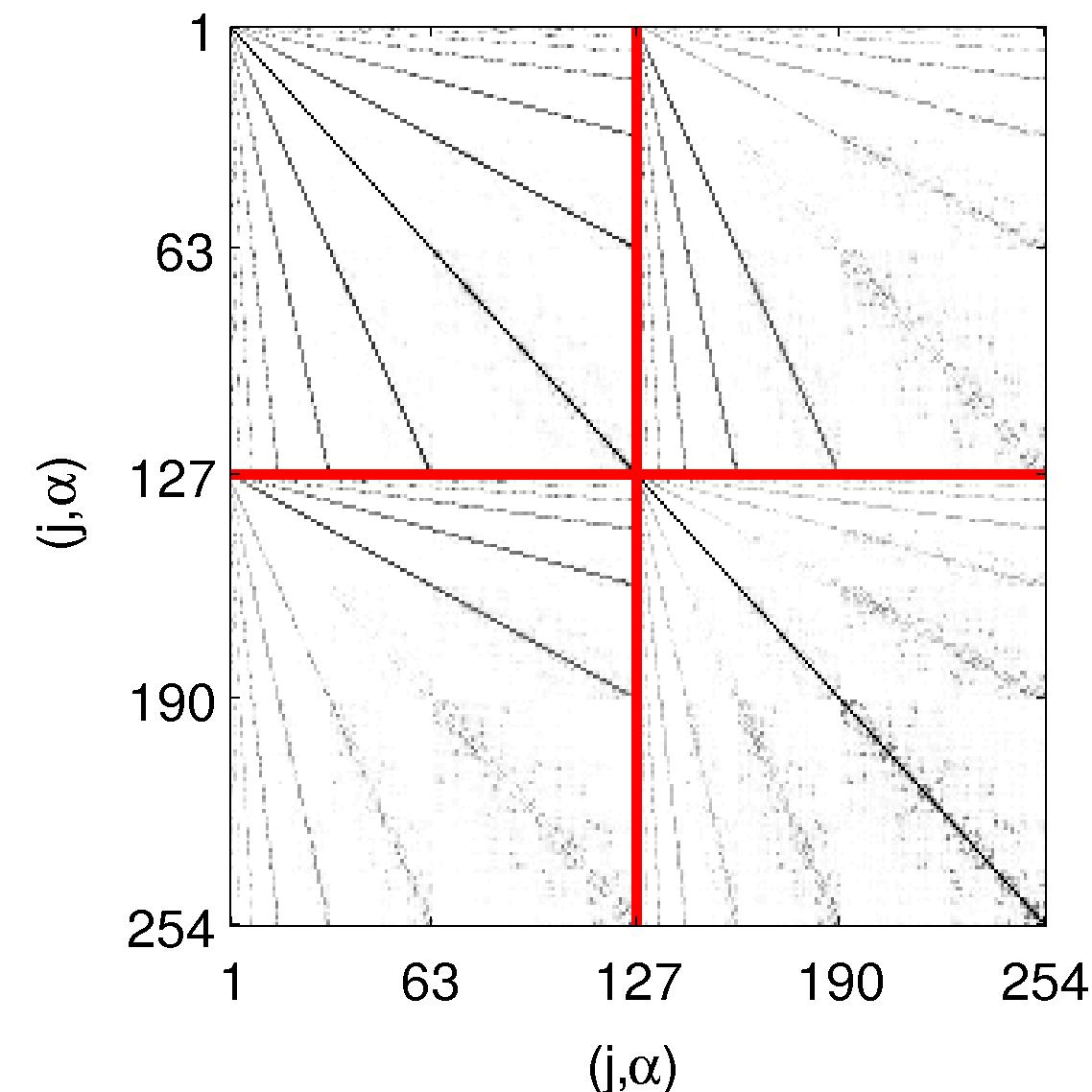}}
\end{subfigure}
\caption[Magnitudes of pairwise inner products of dictionaries
  $\cD_{\textrm{\small Africa},36,1}$ and~$\cD_{\textrm{\small Africa},36,2}$]%
  {\label{fig:sleptrip}Magnitudes of pairwise inner products of
    dictionaries (a)~$\cD_{\textrm{\small Africa},36,1}$ and~(b)~$\cD_{\textrm{\small Africa},36,2}$.
    In (b), thick lines separate inner products
    between the 127 elements with $\alpha=1$ (\textit{top left}) and $\alpha=2$
    (\textit{bottom right}), and their cross products.
    Values are between 0 (white) and 1 (black).
  }
\end{figure}

\begin{figure}[h]
\centering
\includegraphics[width=.485\textwidth]{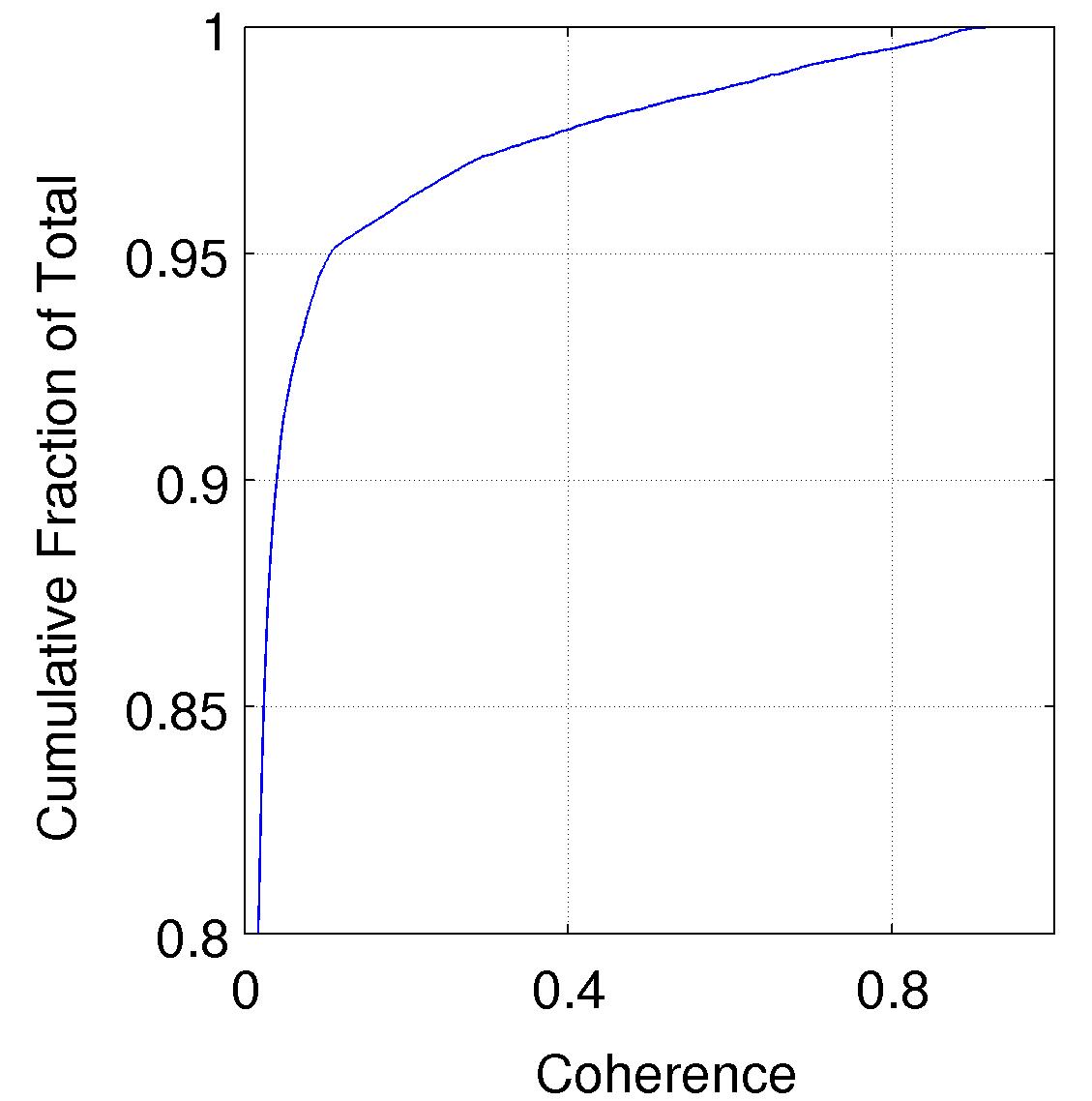}
\includegraphics[width=.485\textwidth]{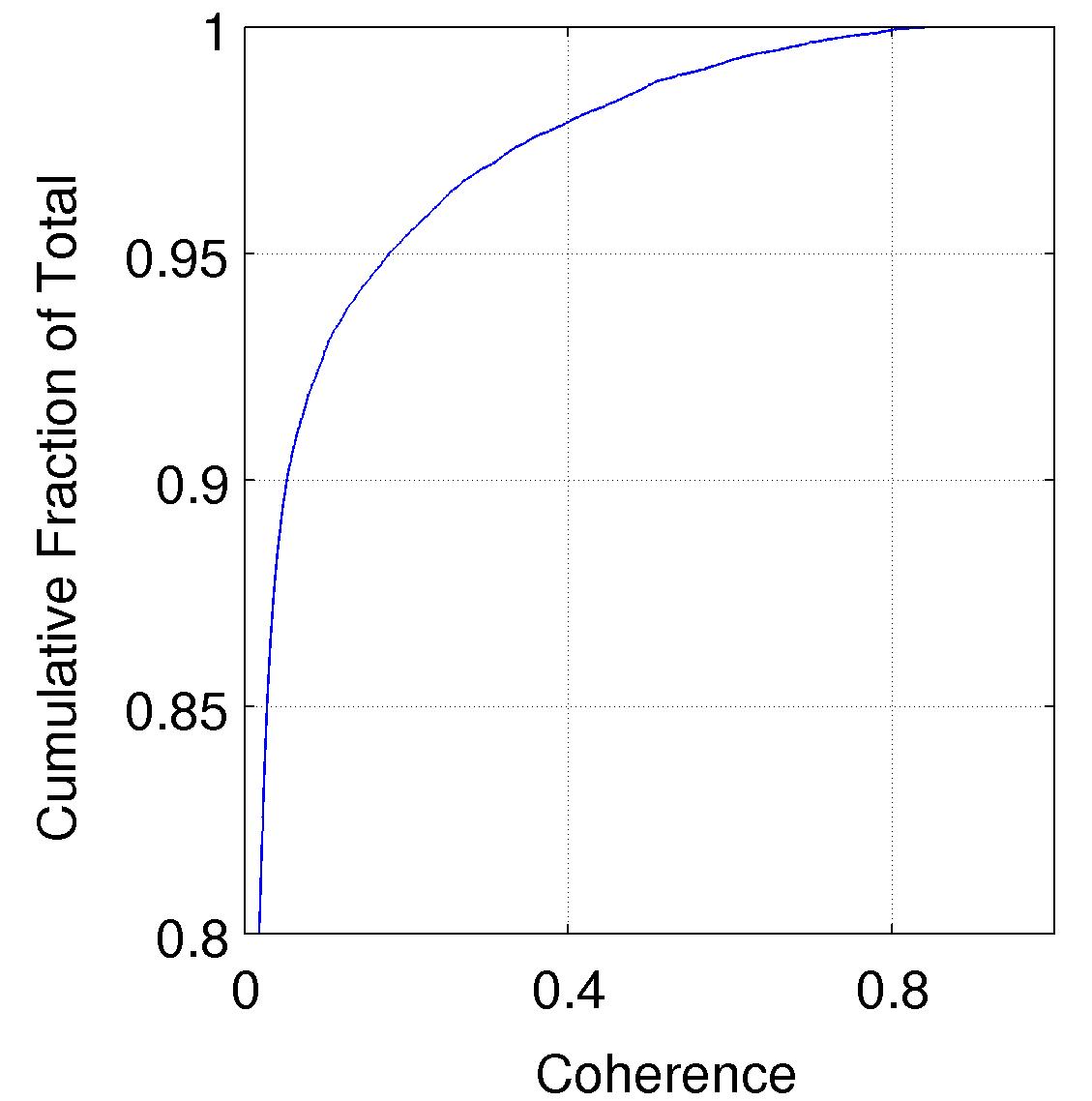}
\caption{\label{fig:sleptripecdf}Empirical
  cumulative distribution functions of pairwise inner product
  magnitudes for dictionaries $\cD_{\textrm{\small Africa},36,1}$ (\textit{left}) and
  $\cD_{\textrm{\small Africa},36,2}$ (\textit{right}).  In both cases,
  approximately $95\%$ of pairwise inner products have a value within $\pm 0.1$.}
\end{figure}

The requirement that the dictionary elements form an orthogonal basis
is less important than the property of mutual incoherence. This we
identify when the inner product between pairs of elements is almost
always very low.  Figure~\ref{fig:sleptrip} shows that the two tree
constructions on continental Africa have good incoherency properties:
most dictionary element pairs are nearly orthogonal.  As can be seen
from Figure~\ref{fig:sleptrip}a, most pairwise inner products are
nearly zero.  More specifically, as expected, dictionary elements
${(j,1)}$, ${(2j,1)}$, ${(2j+1,1)}$, ${(2(2j),1)}$, ${(2(2j)+1,1)}$,
${(2(2j+1), 1)}$, ${(2(2j+1)+1,1), \ldots}$, tend to have large inner
products, while those with non-overlapping borders do not.  This exact
property is also visible in the two diagonal submatrices of
Figure~\ref{fig:sleptrip}b.  In the off-diagonals, due to the
orthonormality of the construction, elements of the form $(j,1)$ and
$(j,2)$ are orthogonal.  In contrast, due to the nature of the tree
subdivision scheme, elements of the form $(2j,1)$ or $(2j+1,1)$ and
$(j,2)$ have a large inner product.  However, the number of
connections between nodes and their ancestors is $\mathcal{O}(n\, [2^H
H])$, while the total number of pairwise inner products is
$\mathcal{O}([n 2^{H}]^2)$; and for reasonably sized values of $L$ the
ratio of ancestral connections to pairwise inner products gets to be
small, see Figure~\ref{fig:sleptripecdf}.

\section{Conclusions}

We have introduced two classes of spherical parameterizations and
discussed their properties in the context of geophysical model
analysis and representation. The first involved the porting of
traditional Cartesian wavelet transforms to the sphere via a
cubed-sphere mapping, with and without special consideration for the
boundaries between the six chunks constituting the entire spherical
surface. The second was a novel elaboration of the classical ideas of
signal concentration on the sphere. Starting from optimally spatially
concentrated bandlimited spherical ``Slepian'' functions, we construct
a dictionary of functions occupying a binary tree, where the elements
of the tree are successive levels of Slepian functions calculated for
an increasingly subdivided spatial domain. 

Both the wavelet bases and the Slepian tree frames that we discuss in
this paper are flexible and efficient ways of studying the information
content and spatiospectral structure of geophysical and cosmological
data. But in both cases their utility will also be derived from using
them in the parameterization of geophysical \textit{inverse}
problems. In ref.~\citenum{Simons+2011b} we discuss an iterative
algorithm that solves an inverse problem in seismic tomography while
promoting sparsity in the cubed-sphere wavelet basis in which the
unknown model is expressed. Ref.~\citenum{Brevdo2011} explores
the inverse problem of approximating bandlimited, heterogeneously
concentrated, essentially multiscale  functions on the sphere from
their samples. 

In this paper and elsewhere, we have shown by example that many
geophysical signals are sparse in the wavelet and Slepian (tree)
``bases''. To tackle large-scale and potentially
ill-conditioned inverse problems, the sparsity of the solution should
actively be encouraged. In either case the inverse problem comes down
to minimizing a mixed 
$\ell_2$-$\ell_1$ functional of the form
\be\label{l1functional2} 
\mF(\bw)=\|\bK\cdot\bS\cdot\bw-\bd\|^2_2+2\lambda
\|\bw\|_1,
\ee
whereby $\bd$ is a data set, $\bK$ a linear operator, and $\bS$ a
certain synthesis map that transforms the set of unknown coefficients
$\bw$ into the domain of $\bK$. Thus, $\bS$ could relate to the
wavelet bases introduced in Section~\ref{firstcon}  or to the
Slepian-tree dictionary introduced in Section~\ref{seccon}. Numerical
experiments conducted for a variety of experimental setups under
geophysically realistic conditions have already indicated that both
types of constructions will be amenable to solving inverse problems in
geophysics and beyond. Further research will
be reported elsewhere.

\section{Acknowledgments}

We thank Jean Charl\'ety, Huub Douma, Massimo Fornasier, Guust Nolet,
Phil Vetter, C\'edric Vonesch and Sergey Voronin for valuable
discussions throughout the past several years. FJS was supported by
Princeton University account 195-2142 and by NSF grant EAR-1014606 to
FJS. Portions of this research were supported by VUB-GOA grant 062 to
ICD and IL, the FWO-Vlaanderen grant G.0564.09N to ICD and IL, and by
NSF grant CMG-0530865 to ICD and others. IL. is Research Associate of
the F.~R.~S.-FNRS (Belgium).

\newpage

%%%%%%%%%%%%%%%%%%%%%%%%%%%%%%%%%%%%%%%%%%%%%%%%%%%%%%%%%%%%%
%%%%% References %%%%%
\bibliography{/home/fjsimons/BIBLIO/bib}
\bibliographystyle{spiebib}

\newpage

\end{document}